\documentclass[11pt]{article}

\usepackage{booktabs} 

\usepackage{latexsym}
\usepackage{epsfig}
\usepackage[mathscr]{eucal}
\usepackage{amsfonts}
\usepackage{amscd}
\usepackage{cite}
\usepackage{array}
\usepackage{bbold}
\usepackage{amssymb}
\usepackage{colordvi}
\usepackage[centertags]{amsmath}
\usepackage{enumerate}
\usepackage{graphicx}
\usepackage{booktabs}
\usepackage{theorem}
\usepackage[footnotesize]{caption}
\usepackage{soul}
\usepackage{url}
\usepackage[hidelinks]{hyperref}
\usepackage{mcite}
\usepackage{slashed}
\usepackage[dvipsnames]{xcolor}
\usepackage{bbm}
\usepackage{placeins}
\usepackage[utf8]{inputenc}
\usepackage{tikz}
\usepackage{tikz-feynman}
\usepackage{physics}
\usepackage{xcolor}
\usepackage{bbm}
\usepackage[utf8]{inputenc}
\usepackage{scrextend}
\usepackage{listings}
\usepackage{multirow}
\usepackage{microtype}

\usepackage{cancel}
\usepackage[normalem]{ulem}
\usepackage[dvipsnames]{xcolor}
\usepackage[absolute,overlay]{textpos}
\usepackage{eso-pic}
\definecolor{codegreen}{rgb}{0,0.6,0}
\definecolor{codegray}{rgb}{0.5,0.5,0.5}
\definecolor{codepurple}{rgb}{0.58,0,0.82}
\definecolor{backcolour}{rgb}{0.95,0.95,0.92}

\definecolor{codegreen}{rgb}{0,0.6,0}
\definecolor{codegray}{rgb}{0.5,0.5,0.5}
\definecolor{codepurple}{rgb}{0.58,0,0.82}
\definecolor{backcolour}{rgb}{0.95,0.95,0.92}
\lstdefinestyle{cppstyle}{
    language=C++,
    backgroundcolor=\color{backcolour},   
    commentstyle=\color{codegreen},
    keywordstyle=\color{blue},
    numberstyle=\tiny\color{codegray},
    stringstyle=\color{codepurple},
    basicstyle=\ttfamily\footnotesize,
    breakatwhitespace=false,         
    breaklines=true,                 
    captionpos=b,                    
    keepspaces=true,                 
    numbers=left,                    
    numbersep=5pt,                  
    showspaces=false,                
    showstringspaces=false,
    showtabs=false,                  
    tabsize=2
}
\definecolor{terminalbg}{RGB}{10,30,60}    
\definecolor{terminalfg}{RGB}{220,230,255} 

\lstdefinelanguage{bash}{
  keywords={cd, ls, cat, echo, export, unset, pwd, mkdir, rm, cp, mv},
  sensitive=true,
  morecomment=[l]{\#},
  morestring=[b]"
}

\lstdefinestyle{bashstyle}{
    language=bash,
    backgroundcolor=\color{black},
    basicstyle=\ttfamily\footnotesize\color{white},
    keywordstyle=\color{cyan},
    commentstyle=\color{green!70!black},
    stringstyle=\color{yellow!80!white},
    numbers=none,
    breaklines=true,
    keepspaces=true,
    breakatwhitespace=false,  
    columns=fullflexible,
    showstringspaces=false,
    breakatwhitespace=false,         
    breaklines=true,                 
    captionpos=b,                    
    keepspaces=true,                 
    numbers=left,                    
    numbersep=5pt,                  
    showspaces=false,                
    showstringspaces=false,
    showtabs=false,                  
    tabsize=2
}

\lstdefinestyle{csvstyle}{
    style=cppstyle,
    alsoletter={,},                
    moredelim=[s][\color{gray}]{,}{,} 
}
\allowdisplaybreaks{}

\numberwithin{equation}{section}

\usepackage[
  a4paper,
  left=1cm,
  right=1cm,
  top=2.5cm,
  bottom=2.5cm,
  headheight=0pt,
  headsep=0pt
]{geometry}

\newcommand{\s}{\newline \vspace*{-3.5mm}}
\newcommand{\be}{\begin{equation}}
\newcommand{\ee}{\end{equation}}
\newcommand{\beq}{\begin{eqnarray}}
\newcommand{\eeq}{\end{eqnarray}}

\newcommand{\lsim}{\raisebox{-0.13cm}{~\shortstack{$<$ \\[-0.07cm]
      $\sim$}}~}

\newcommand{\RNum}[1]{\uppercase\expandafter{\romannumeral #1\relax}}

\newcommand{\si}[1]{\text}
\newcommand{\SI}[1]{\text}

\newcommand{\ttt}[1]{\texttt{#1}}

\newcommand{\MM}[1]{{\color{magenta}MM: #1}}
\newcommand{\PB}[1]{{\color{blue}PB: #1}}
\newcommand{\KE}[1]{{\color{red}KE: #1}}
\newcommand{\RS}[1]{\color{orange}RS: #1}
\newcommand{\PG}[1]{\textbf{\color{ForestGreen}#1}}

\newcommand{\rotation}[0]{\left( \begin{array}{cc} \cos \alpha & \sin \alpha \\ - \sin\alpha & \cos \alpha \end{array} \right) }

\newcommand{\Z}[1]{\ensuremath{\mathbbm{Z}_{#1}}} 

\definecolor{darkgreen}{rgb}{0.13, 0.50, 0.09}

\newcommand\refeq[1]{Eq.~(\ref{#1})}
\def\reffi#1{\mbox{Fig.~\ref{#1}}}

\newcommand{\ct}[1]{\delta #1}
\newcommand{\ctZ}[1]{\delta Z_{#1}}
\newcommand{\rse}[2]{\hat{\Sigma}_{#1}(#2)}
\newcommand{\rseL}[2]{\hat{\Sigma}_{#1}^L(#2)}
\newcommand{\rseR}[2]{\hat{\Sigma}_{#1}^R(#2)}
\newcommand{\rseS}[2]{\hat{\Sigma}_{#1}^S(#2)}
\newcommand{\se}[3]{\Sigma_{#1}^{#3}(#2)}

\newcommand{\pderiv}[2]{\frac{\partial #1}{\partial #2}}

\AddToShipoutPictureFG*{%
  \AtPageUpperLeft{%
    \hspace*{\dimexpr\paperwidth-6cm\relax}%
    \raisebox{-1cm}{%
      \parbox[t]{5cm}{%
        \raggedleft
        \small
        KA-TP-16-2026\\
      }%
    }%
  }%
}
\begin{document}

\title{Full Next-To Leading-Order Electroweak and QCD Corrections \\ to the Relic Density in the CxSM}
\author{
Pavao Brica$^{1\,}$\footnote{E-mail: \texttt{pavao.brica@partner.kit.edu}},
Karim Elyaouti$^{1\,}$\footnote{E-mail: \texttt{karim.elyaouti@partner.kit.edu}},
Pedro Gabriel$^{1,2\,}$\footnote{E-mail: \texttt{ptgabriel@fc.ul.pt}},\\
Margarete M\"{u}hlleitner$^{1\,}$\footnote{E-mail: \texttt{margarete.muehlleitner@kit.edu}},
Rui Santos$^{2,3\,}$\footnote{E-mail: \texttt{rasantos@fc.ul.pt}}
\\[9mm]
{\small\it
$^1$Institute for Theoretical Physics, Karlsruhe Institute of Technology,} \\
{\small\it Wolfgang-Gaede-Str. 1, 76131 Karlsruhe, Germany}\\[3mm]
{\small\it $^2$Centro de F\'{\i}sica Te\'{o}rica e Computacional,
    Faculdade de Ci\^{e}ncias,} \\
{\small \it    Universidade de Lisboa, Campo Grande, Edif\'{\i}cio C8
  1749-016 Lisboa, Portugal} \\[3mm]
{\small\it
$^3$Departamento de F\'{\i}sica, Faculdade de Ci\^{e}ncias,} \\
{\small \it   Universidade de Lisboa, 1749-016 Lisboa, Portugal} \\[3mm]
 %
}

\maketitle

\abstract{In this work we present the calculation of the next-to-leading order (NLO) QCD and electroweak corrections to the thermally averaged cross section of Dark Matter (DM) annihilation in the framework of the complex singlet extended Standard Model (CxSM) with one DM candidate. We derive the corresponding NLO QCD and electroweak corrected relic density and discuss the impact of the corrections as well as the remaining theoretical uncertainty due to missing higher-order corrections. This is estimated through a variation of the renormalization schemes for the singlet vacuum expectation value $v_S$ and the Higgs mixing angle $\alpha$. For the bulk of the scanned parameter points, the QCD and electroweak corrections are found to be of typical size with the remaining theoretical uncertainty ranging at the percent level. The larger corrections that are found, can be attributed to a large counterterm for $v_S$ emerging in the case of large mass gaps in the decay channel used for the renormalization. 
The corrections are of phenomenological impact. There are parameter points that are allowed at leading order (LO), but are  excluded after including the NLO corrections to the relic density, and vice versa. 
Our corrections have been implemented in the new code \texttt{RelExt@NLO} based on the LO code \texttt{RelExt}, which has been made publicly available. This work marks a first step towards the calculation of the QCD and electroweak corrected relic densities in models with DM candidates stabilized by a discrete $\mathbb{Z}_2$ symmetry.}
\newpage
\tableofcontents

\newpage


\section{Introduction}
\label{sec:Introduction}

The SM has been tested by numerous experiments to the highest accuracy and proven to be very successful in predicting experimental observables. There are, however, phenomena that cannot be explained by the SM and that remain to be solved. Among these, Dark Matter (DM) is probably one of the greatest mysteries of science today. The SM does not contain any particle that can explain the observations quantitatively. A plethora of theories beyond the SM hence suggest many different DM candidates with masses spanning a huge range from ultra-light scalars or axion(-like) particles up to solar mass objects. Any of these models has to make sure that the generated DM relic density at least does not exceed the measured value and that the DM candidates are compatible with the direct and indirect DM experiments. \s

In this work, we assume DM to have particle character and to be non-relativistic. We furthermore assume that today's observed DM abundance  of $\Omega_{\text{DM}} h^2 = 0.1200 \pm 0.0012$ \cite{Planck:2018vyg} was generated thermally through freeze-out. We consider a Higgs portal model, where a discrete symmetry ensures the stability of the DM candidate(s). Higgs bosons that couple to both the SM and the dark sector particles provide the link to the visible world. For the computation of the relic density, the Boltzmann equation, which describes the evolution of the DM density from thermal equilibrium until today, has to be solved. It crucially depends on the thermally averaged cross section (TAC) describing annihilation or co-annihilation of DM particles into SM particles. Being inversely proportional to the TAC, the relic density decreases for efficient annihilation and hence strong coupling between the dark sector particles and the Higgs portal. On the other hand, such a strong link may lead to DM-nucleon scattering cross sections that exceed current experimental limits on DM direct detection. There is hence a tension between the compatibility with the constraints from the relic density and direct detection. \s

In order to be able to identify the true model underlying nature and delineate its still allowed parameter space, we have to make precise theory predictions for derived input parameters and observables that are compared with the experimental results, which are becoming more and more precise. 
Although there exist many codes that compute the DM relic density at leading order (LO), for a review cf.~e.g.~\cite{Arina:2020alg}, this is not the case when it comes to the inclusion of higher-order corrections. The code \texttt{MicrOMEGAs} \cite{Belanger:2001fz,Belanger:2006is} provides predefined functions to improve the TAC to NLO accuracy, and the code \texttt{DM@NLO} \cite{Harz:2023llw} provides the next-to-leading order (NLO) (supersymmetric) QCD corrections to DM (co-)annihilation in the Minimal Supersymmetric extension of the SM (MSSM). In \cite{Banerjee:2019luv,Banerjee:2021xdp,Banerjee:2021hal,Banerjee:2021oxc,Banerjee:2021anv}, the NLO corrections to the thermally averaged cross section were provided for the Inert Doublet Model.
\s 

Recently, we published the \texttt{C++} code \texttt{RelExt} \cite{Capucha:2025iml} which allows to efficiently scan the parameter spaces for arbitrary models with DM candidates that are stabilised by a discrete $\mathbb{Z}_2$ symmetry with the goal to find parameter combinations that lead to the correct relic density. In this work, we take the code to the next level of precision, by computing the NLO QCD and electroweak (EW) corrections to the DM annihilation processes into SM pairs that enter the relic density through the TAC. We will perform the calculation in the framework of the complex singlet extended SM (CxSM). This model with one DM candidate and two visible Higgs bosons will serve as a first rather simple sample case to the later envisioned extension of \texttt{RelExt} to the precision computation of the relic density at NLO for arbitrary models with DM candidates stabilised by discrete $\mathbb{Z}_2$ symmetries. 
Our corrections are implemented in the existing tool \texttt{RelExt}. The new code \texttt{RelExt}@\texttt{NLO} can be downloaded from the url:

\centerline{\url{https://github.com/jplotnikov99/RelExt/tree/NLO}} 
\medskip
The paper is organized as follows. In Sec.~\ref{sec:model}, we introduce the model. In Sec.~\ref{sec:nloreldec}, we give the formula for the NLO relic density. Section \ref{sec:Renormalization} is devoted to the description of our renormalizaton conditions. The treatment of the infared divergences is outlined in Sec.~\ref{sec:IRDivergencies}. In Sec.~\ref{sec:progdescr} we give a small program description. Section \ref{sec:param} describes our parameter scan. Section \ref{sec:numericalanalysis} contains our numerical analysis, where we present the overall size of the corrections, discuss the relative size and the impact of the QCD and EW corrections on the TAC and on the relic density, and we give an uncertainty estimate due to missing higher-order corrections derived from changing consistently the renormalization scheme. Our conclusions are given in Sec.~\ref{sec:conclusions}.


\section{The CxSM}
\label{sec:model}

In the CxSM \cite{Costa:2015llh,Egle_2022}, the SM is extended by an additional complex scalar singlet field $\mathbb{S}$ with zero isospin and hypercharge. After electroweak symmetry breaking, the doublet field $\Phi$ and the singlet field $\mathbb{S}$ can be parametrized as 
\begin{eqnarray}
\Phi = \left( \begin{array}{c}
G^+ \\ \frac{1}{\sqrt{2}} (v + H + iG^0) \end{array}
\right) \,, \quad \mathbb{S} = \frac{1}{\sqrt{2}} (v_S + S + i (v_A + A)).\label{eq: cxsm_content}
\end{eqnarray}
Here, $H$, $S$ and $A$ are real scalar fields and $G^+$ and $G^0$ are the charged and neutral Goldstone bosons, respectively. The vacuum expectation values (VEVs) of the neutral CP-even Higgs $H$, the CP-even singlet field $S$, and the CP-odd singlet field $A$ are given by $v$, $v_S$, and $v_A$, respectively. To ensure the existence of a stable DM candidate, we require that the Lagrangian is invariant under two separate $\Z{2}$ symmetries. The transformations of the singlet fields $S$ and $A$ under these symmetries read
\begin{equation}
    A\rightarrow-A \quad \mbox{ and } \quad S\rightarrow-S \;.
\end{equation}
We furthermore impose a global $U(1)$ symmetry on the scalar singlet that is softly broken.
The corresponding renormalizable potential reads 
\begin{eqnarray}\label{eq::CxSMPotential}
V = \frac{m^2}{2} |\Phi|^2 + \frac{\lambda}{4} |\Phi|^4 + \frac{\delta_2}{2} |\Phi|^2 |\mathbb{S}|^2 + \frac{b_2}{2} |\mathbb{S}|^2 + \frac{d_2}{4} |\mathbb{S}|^4 + \left( \frac{b_1}{4} \mathbb{S}^2 + c.c. \right) \;,
\end{eqnarray}
where all masses and couplings are real. The soft breaking is induced by the non-zero value of $b_1$. In this work, we consider $v_A = 0$, which implies that the $A\rightarrow-A$ symmetry remains unbroken. Therefore, $A$ is stable and  becomes the DM candidate of the model. As $v_S$ is chosen to be non-zero, the other $\mathbb{Z}_2$ symmetry is broken and the scalar fields $H$ and $S$ mix. The mass eigenstates $h_i$ ($i=1,2$) are obtained by rotating the gauge eigenstates $H$ and $S$ through a rotation matrix $R_\alpha$, which reads
\begin{equation}
    R_\alpha = \rotation \;. 
    \label{eq:rot_matrix}
\end{equation}
The mass eigenstates are given by
\begin{equation}
    \left(\begin{array}{c}
    h_1 \\
    h_2 \end{array}\right) = R_\alpha \left(\begin{array}{c}
    H \\
    S \end{array}\right) \; .
    \label{eq:field_rot_CPEven}
\end{equation}
The mass matrix of the gauge eigenstates $\mathcal{M}$ reads 
\begin{equation}
    \mathcal{M} = \left(\begin{array}{cc}
    \frac{v^2\lambda}{2} & \frac{\delta_2vv_S}{2}\\
    \frac{\delta_2vv_S}{2}&\frac{d_2v_S^2}{2} \end{array}\right) + \left(\begin{array}{cc}
    T_1 & 0\\
    0 & T_2 \end{array}\right)\; ,
    \label{eq:mass_matrix}
\end{equation}
where the tadpole parameters $T_i$ $(i=1,2)$ are obtained from the minimization conditions as
\begin{subequations}
\begin{align}
    \frac{\partial V}{\partial v} &\equiv T_1 \Rightarrow \frac{T_1}{v} = \frac{m^2}{2} + \frac{\delta_2v_S^2}{4} + \frac{v^2\lambda}{4} \; ,\\
    \frac{\partial V}{\partial v_S} &\equiv T_2 \Rightarrow \frac{T_2}{v_S} = \frac{b_1+b_2}{2} + \frac{\delta_2v^2}{4} + \frac{v_S^2d_2}{4} \; .
\end{align}
\label{eq:tadpoles}
\end{subequations}
They are zero at tree level, but introduced here already as they can become non-zero at higher orders. The mass of the DM  candidate $A$ is given by
\begin{equation}
    m_A^2 = \frac{-b_1+b_2}{2} + \frac{\delta_2v^2}{4} + \frac{v_S^2d_2}{4} = -b_1 + \frac{T_2}{v_S}\; .
\end{equation}
The masses of the mass eigenstates $h_i$ are given by the eigenvalues of the mass matrix $\mathcal{M}$,
\begin{eqnarray}
    D^2_{hh} = R_\alpha \mathcal{M}R^T_\alpha, \quad D^2_{hh} = \mathrm{diag}(m^2_{h_1}, m^2_{h_2})\;,
\end{eqnarray}
with the mass ordering $m_{h_1} < m_{h_2}$. Due to the mixing of the two visible scalar particles, their coupling constants to the SM particles are modified by an additional factor $\kappa_i$ w.r.t.~the corresponding SM coupling, 
\begin{eqnarray}
    g_\mathrm{h_i SMSM} = g_\mathrm{h_{SM}SMSM} \kappa_i,\; \kappa_i = \begin{cases}
    \cos{\alpha}, \; i=1\\
    \sin{\alpha}, \; i=2 
    \end{cases} \, .
\end{eqnarray}
Here, $h_\mathrm{SM}$ denotes the SM Higgs boson, and $g_\mathrm{h_{SM}SMSM}$ the corresponding coupling of the SM Higgs boson to a pair of SM particles.
The chosen input parameter set is given by
\begin{equation}
    v, \; v_S, \; \alpha, \; m_{h_1}, \; m_{h_2}, \; m_A \; .
\end{equation}


\section{The NLO Relic Density \label{sec:nloreldec}}
In general, the dark sector of beyond-SM (BSM) extensions can consist of $N$ particles $\chi_i$ ($i= 1,\dots,N$) that are charged under a new dark quantum number. The lightest of these particles, $\chi_1$ with mass $m_1$, corresponds to the DM candidate, as all other dark sector particles $\chi_j$ ($j=2,\dots,N$) will eventually decay into this particle. 
The relic density can be calculated as \cite{Belanger:2006is}
\begin{equation}
\Omega_\text{DM}h^2=2.742\cdot10^8\frac{m_1}{\text{GeV}}Y(x_0)\,,
\end{equation}
where $x_0=m_1/T_0$ in terms of the DM mass $m_1$ and the temperature $T_0$ of the universe today. The yield $Y$ is obtained from the solution of the Boltzmann equation from $x_0$ to $x=m_1/T$, where $T$ denotes the temperature of the early universe \cite{Edsjo:1997bg}, 
\begin{equation}\label{eq: boltz eq}
 \frac{\mathrm{d}Y}{\mathrm{d}x} = - \sqrt{\frac{\pi}{45G}}\frac{g^{1/2}_*}{x^2}\langle\sigma v\rangle_\text{eff} \left(Y^2 - Y_\text{eq}^2\right) \, ,
\end{equation}
$G$ is the gravitational constant and 
\begin{equation}
    g^{1/2}_* = \frac{h_\text{eff}}{\sqrt{g_\text{eff}}}\left (1+\frac{T}{3h_\text{eff}}\frac{\mathrm{d}h_\text{eff}}{\mathrm{d}T}\right) \,,
\end{equation}
in terms of the effective degrees of freedom $h_\text{eff}$ and $g_\text{eff}$ of the entropy density and the energy density, respectively. The equilibrium yield $Y_\text{eq}$ in the non-relativistic limit is given by
\begin{equation}
    Y_\text{eq} = \sum_i Y_{i,\text{eq}} = \frac{45 x^2}{4\pi^4 h_\text{eff}(x)}\sum^N_{i} g_i \cdot\left(\frac{m_i}{m_1} \right)^2 \cdot K_2\left(\frac{m_i}{m_1}x\right) \,,
\end{equation}
with the sum performed over all dark sector particles $i$ and  $m_i$ denoting their corresponding mass. The TAC $\langle\sigma v\rangle_\text{eff} $ for the (co-)annihilation of the dark sector particles $i$ and $j$ into the thermal bath particles $k$ and $l$, is given by
\begin{equation}\label{eq:TAC}
    \langle\sigma v\rangle_\text{eff} = \frac{\sum_{i,j=1}^N g_i g_j \int_{s_\text{min}}^\infty \mathrm{d}s \sqrt{s}p_{ij}^2 \sigma_{ij}(s) K_1\left(\frac{\sqrt{s}x}{m_1}\right)}{2T\left(\sum_{i=1}^N g_i m_i^2K_2\left(\frac{m_i}{m_1}x\right)\right)^2} \,,
\end{equation}
where the value of the lower integration bound $s_\text{min}$ is determined by the larger of the kinematic thresholds for the initial and final states. If the final-state particles are heavier than the initial-state particles, the center-of-mass energy must be at least equal to the combined rest-mass of the final-state particles in order for the process to be kinematically allowed. Hence,
\begin{eqnarray}
    s_\text{min} = \mathrm{max}((m_i+m_j)^2,(m_k+m_l)^2)\;.
\end{eqnarray}
The $K_n$ ($n=1,2$) are the modified Bessel functions of the second kind at order $n$, $g_i$ and $g_j$ are the internal degrees of freedom of the dark sector particles and $\sqrt{s}$ is the center-of-mass energy of the (co-)annihilation process. Furthermore, $p_{ij}$ is given by
\begin{equation}
    p_{ij} = \frac{\sqrt{(s-(m_i+m_j)^2)(s-(m_i-m_j)^2)}}{2\sqrt{s}} \,.
\end{equation} 
The cross section $\sigma_{ij}$ denotes the cross section for the (co-)annihilation of two dark sector particles $i,j$ into all possible freeze-out final states. In the following, we will only consider two-body final states $k$ and $l$, and we need not consider co-annihilation as we have only one DM particle, such that at LO the cross section is given by
\begin{equation}\label{eq:siglo}\sigma^{\text{LO}}_{ii} = \frac{1}{32\pi sg_ig_j} \sum_{k,l} \frac{\lambda^{1/2}(s,m_k^2,m_l^2)}{\lambda^{1/2}(s,m_i^2,m_i^2)}\int |\mathcal{M}_{ii\rightarrow kl}^{\text{LO}}|^{2} \dd\cos\theta \,. 
\end{equation}
Here, $\lambda$ is the Käll\'en function, $\theta$ the azimuthal scattering angle, and $\mathcal{M}_{ii\rightarrow kl}^{\text{LO}}$ the LO $2\to 2$ annihilation matrix element. Our goal is to include the NLO QCD and EW corrections to the TAC, i.e.~to compute
\beq
\sigma_{ii}^{\text{corr}}
= \sigma_{ii}^{\text{LO}} + \sigma_{ii}^{\text{NLO}} \;.
 \label{eq:sigmacorr}
\eeq
The NLO-corrected annihilation cross section $\sigma^{\text{corr}}$ is obtained by replacing 
in Eq.~(\ref{eq:siglo}) the LO matrix element squared by the NLO-corrected matrix element squared, 
\beq
|\mathcal{M}_{ii\rightarrow kl}^{\text{corr}}|^2 = |\mathcal{M}_{ii\rightarrow kl}^{\text{LO}}|^2 + 2 \mbox{Re} (\mathcal{M}_{ii\rightarrow kl}^{\text{LO}} \mathcal{M}_{ii\rightarrow kl}^{* \text{NLO}}) \;.
\label{eq:NLO_amp}
\eeq
In case of infrared corrections, we also have to include the real corrections from $2 \to 3$ processes. We will come back to this point later in more detail.
The expression for the real corrections from the $2\to3$  annihilation cross section $\sigma^\text{real}$ is taken from Ref.~\cite{romaothree} and reads
\begin{equation}
    \begin{split}
    \sigma^\text{real} = &\frac{1}{64(2\pi)^4}\frac{1}{\sqrt{(p_a\cdot p_b)^2-m^2_a m^2_b}} \int^{E_{3,\mathrm{max}}}_{m_3}\mathrm{d}E_3\sqrt{E_3-m^2_3}\int^\pi_0\mathrm{d}\theta \sin{\theta}\\&\times\int^\pi_0\mathrm{d}\theta^*\sin{\theta^*}\int^{2\pi}_{0}\mathrm{d}\varphi^*\frac{\lambda(m_{12},m_1,m_2)}{2m^2_{12}}|\mathcal{M}_{2\to3}|^2\;.
\end{split}
\end{equation}
Here, ${\cal M}_{2\to 3}$ denotes the matrix element of the real correction process, $p_{a,b}$ and $m_{a,b}$ are the four-momenta and masses of the two incoming particles $a$ and $b$, respectively. The momenta $p_i$, the masses $m_i$ and the energy $E_i$ correspond to the $i$th final state particle ($i=1,2,3$) and $m_{12}^2 \equiv (p_1+p_2)^2$.

\section{Renormalization}
\label{sec:Renormalization}
The loop integrals in the computation of the NLO corrections can lead to UV-divergent amplitudes that need to be cancelled through the process of renormalization. 
For this, we replace the bare parameters and wave functions by their renormalized counterparts, thus introducing counterterms that cancel the divergent parts of the loop integrals. For the bare parameter $p_0$ we have 
\begin{align}
    p_0&= p+\ct{p},
    \label{eq:renorm_param}
\end{align}    
where $p$ denotes the renormalized parameter and $\delta p$ its counterterm. Accordingly, we replace the bare field $\phi_0$ by 
\begin{align}    
\phi_0&=\sqrt{Z_\phi}\phi\approx\left(1+\frac{\ct{Z_{\phi}}}{2}\right)\phi,
    \label{eq:renorm_wf}
\end{align}
with $\phi$ denoting the renormalized field and $\delta Z_\phi$ the wave function renormalization constant (WFRC).
In the following subsections, we introduce the renormalizaton conditions used to fix the finite parts of the UV-divergent counterterms of the parameters and the WFRCs that contribute to the computation of the NLO QCD and EW corrections to the TAC. For further details on the renormalization conditions, we also refer to Refs.~\cite{Egle:2022wmq,Egle:2023pbm}, where we introduced them for the the first time. 


\subsection{Tadpoles}
\label{subsec:tadpoles}

\begin{figure}
    \centering
    \scalebox{1}{
        
         \begin{tikzpicture}
            \begin{feynman}
                \vertex[blob] (d) at (0,0.5) {};
                \vertex at(0,-1) (c);
                \vertex at (-2,-1) (a)  {\(\text{DM}\)};
                \vertex at (2,-1) (b) {\(\text{DM}\)};

                \diagram* {
                    (a) -- [scalar] (c),
                    (c) -- [scalar] (b),
                    (c) -- [scalar, edge label'=\(h_i\)]  (d),                
                };
            \end{feynman}
        \end{tikzpicture}

        \begin{tikzpicture}
            \begin{feynman}
                \vertex at (-2,0) (a)  {\(h_i\)};
                \vertex at (1.5,1) (b) {\(\text{DM}\)};
                \vertex at (1.5,-1) (c) {\(\text{DM}\)};
                \vertex at (0,0) (d);
                \vertex[blob] (m) at (-0.5,1) {};

                \diagram* {
                    (a) -- [scalar] (d),
                    (d) -- [scalar] (b),
                    (d) -- [scalar] (c),
                    (d) -- [scalar, edge label'=\(h_j\)]  (m),                
                };
            \end{feynman}
        \end{tikzpicture}      
    }
    \caption{Examples of additional diagrams with tadpoles that result from the application of the ATS.}
    \label{fig:tadpole_diagrams}
\end{figure}
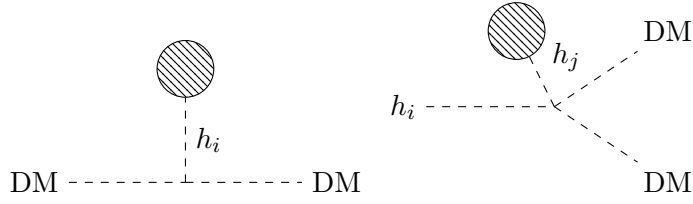

There are different ways to treat the tadpoles. In the Standard Tadpole Scheme (STS) 
the tadpoles defined in \refeq{eq:tadpoles} are taken as inputs. Using the definition \refeq{eq:renorm_param}, we then apply the following renormalization condition, 
\begin{equation}
    T_{i,0}=T_i-\ct{T_i}=0,\quad i=1,2\; .
    \label{eq:STS_condition}
\end{equation}
In the STS, the WFRCs and any counterterm that depends on them are gauge-dependent, however. This can be cured by resorting to the alternative approach suggested in Ref.~\cite{Fleischer:1980ub}. In this so-called Alternative Tadpole Scheme (ATS), we renormalize the scalar VEVs instead of the tadpoles and we require the renormalized VEVs to be the same as the tree-level VEVs. Hence, we get the relations
\begin{align}
    v_{R}&=v+\Delta v\; , \\
    v_{S,R}&=v_s+\Delta v_S\; ,
\end{align}
where the subscript $R$ indicates the renormalized quantity and $v$ and $v_s$ represent the tree-level VEVs. The VEV counterterms can be related to the tadpole counterterms using the mass matrix in \refeq{eq:mass_matrix}, through the expression
\begin{equation}
    \left(
    \begin{array}{c}
        \ct{T_1} \\
        \ct{T_2}  
    \end{array} 
    \right) = \mathcal{M} \left(\begin{array}{c}
        \Delta v  \\
        \Delta v_S 
    \end{array} \right) \; .
\end{equation}
Inverting the equation and rotating the tadpole side to the mass basis we end up with the expression
\begin{equation}
    \left(\begin{array}{c}
        \Delta v  \\
        \Delta v_S 
    \end{array} \right) = R_\alpha^T(D_{hh}^2)^{-1}
    \left(
    \begin{array}{c}
        \ct{T_{h_1}} \\
        \ct{T_{h_2}}  
    \end{array} 
    \right)=
    R_\alpha^T
    \left(
    \begin{array}{c}
        \frac{T_{h_1}}{m_{h_1}^2} \\
        \frac{T_{h_2}}{m_{h_2}^2} 
    \end{array} 
    \right)\; ,
    \label{eq:vev_TP_relation}
\end{equation}
where we used the renormalization condition in \refeq{eq:STS_condition} in the last step. Rewriting Eq.~(\ref{eq:vev_TP_relation}) as, 
\begin{equation}
    \left(\begin{array}{c}
        \Delta v  \\
        \Delta v_S 
    \end{array} \right) =
    R(\alpha)
    \left(
    \begin{array}{c}
        \frac{-i}{m_{h_1}^2}iT_{h_1} \\
        \frac{-i}{m_{h_2}^2}iT_{h_2} 
    \end{array} 
    \right)\; ,
\end{equation}
the quantities ($i=1,2$)
\begin{eqnarray}
\Delta v_i \equiv  \frac{-i}{m_{h_i}^2} i T_{h_i}
\end{eqnarray}
can be interpreted as connected tadpole diagrams, containing the Higgs tadpole and its propagator at zero momentum transfer \cite{Krause:2017mal}.
This means that in the ATS all tree-level vertices that depend on $v$ or $v_S$, will generate at NLO additional tadpole diagrams like the ones represented in \reffi{fig:tadpole_diagrams}. \s

Since we take the VEVs as input parameters, after applying the ATS we still have two counter-terms for the bare parameters $v_0$ and $v_{S,0}$, respectively. In accordance with \refeq{eq:renorm_param}, we get the expressions
\begin{align}
    v_0&=v_R+\ct{v} \; ,\\
    v_{S,0}&=v_{S,R}+\ct{v_S} \; ,
\end{align}
where we added the index $R$ to denote the renormalized VEVs. 
From now on, we will drop the index $R$. To fix $\ct{v}$ we can use the known relation to the SM parameters,
\begin{equation}
    v=\frac{2 m_W\sqrt{m_Z^2-m_W^2}}{e m_Z}\; ,
\end{equation}
where $e$ denotes the electric charge and $m_W$ and $m_Z$ the $W$ and $Z$ boson mass, respectively. This allows us to express $\ct{v}$ as a function of the counterterms for $m_W^2$, $m_Z^2$ and the electric charge, such that
\begin{equation}
    \ct{v}=v\left[\frac{1}{2\left(m_Z^2-m_W^2\right)}\left(\frac{m_Z^2}{m_W^2}\ct{m_W^2}-\frac{m_W^2}{m_Z^2}\ct{m_Z^2}\right)+\ct{Z_e}\right]\; ,
    \label{eq:CT_vev_dependence}
\end{equation}
where we used the relation $\ct{e}=e \ct{Z_e}$. The renormalization conditions for fixing these counterterms will be discussed in the following subsections. \s

In the case of $\ct{v_S}$ there are no known relations with other parameters and therefore it must be fixed differently. This will be discussed in subsection \ref{subsec:renorm_vs}.


\subsection{Masses and Wave Functions}
\label{subsec:masses_wf}
For the renormalization of the mass parameters and the wave functions we choose the On-Shell (OS) scheme as described e.g.~in \cite{Denner:1991kt}. In this physically motivated scheme, the renormalized two-point functions, i.e.~propagators, are required to have poles at the physical masses $m$ with unit residue. We introduce the renormalized propagator correction, the self-energy $\hat{\Sigma} (p^2)$, for the scalar field $\phi$ with four-momentum $p$ as
\begin{equation}
    \rse{\phi}{p^2}=\se{\phi}{p^2}{\text{tad}}-\ct{m^2}+\ct{Z_{\phi}}(p^2-m^2) \;,
    \label{eq:renorm_SE}
\end{equation}
where $\se{\phi}{p^2}{\text{tad}}$ represents the one-loop self-energy correction including the tadpole contributions (cf.~Fig.~\ref{fig:tadpole_diagrams} (left)) as required in the ATS, $\delta m^2$ is the mass counterterm and $\delta Z_{\phi}$ the WFRC. 
The mass counterterm in the OS scheme is given by 
\begin{equation}
    \ct{m^2}=\se{\phi}{m^2}{
    \text{tad}} \;,
    \label{eq:dm}
\end{equation}
%
and the WFRC is fixed through the condition
\begin{equation}
    \ct{Z_{\phi}}=-\pderiv{\se{\phi}{p^2}{\text{tad}}}{p^2}\bigg|_{p^2=m^2} \;.
    \label{eq:dZ}
\end{equation}
Fields with the same quantum numbers can mix at one-loop level so that the structure of the WFRCs has to be adjusted accordingly, which now takes a matrix form. The generalisation of \refeq{eq:renorm_wf} to $n$ mixing fields $\phi_i$ ($i=1,...,n$) leads to
\begin{equation}
    \sqrt{Z_\phi} \approx \mathbb{1}_{n\times n}+\frac{\ctZ{\phi}}{2}=\left(
    \begin{array}{cccc}
        1+\frac{\ctZ{\phi_1\phi_1}}{2} & \frac{\ctZ{\phi_1\phi_2}}{2}   & \cdots & \frac{\ctZ{\phi_1\phi_n}}{2}\\
        \frac{\ctZ{\phi_2\phi_1}}{2}   & 1+\frac{\ctZ{\phi_2\phi_2}}{2} & \;     & \vdots\\
        \vdots                         & \;                             & \ddots & \vdots \\
        \frac{\ctZ{\phi_n\phi_1}}{2}   & \cdots                         &\cdots  &  1+\frac{\ctZ{\phi_n\phi_n}}{2} \\
    \end{array}\right)\; ,
    \label{eq:wfrc_mixing}
\end{equation}
where $\mathbb{1}_{n\times n}$ represents the $n\times n$ unit matrix. 
In order to fix the non-diagonal components of the WFRC matrix, in the OS scheme we require the renormalized one-particle irreducible two-point functions to be diagonal if the external lines are on their mass shell, i.e.~the off-diagonal elements of the renormalized self-energies have to fulfill
\begin{equation}
    \rse{\phi_i\phi_j}{m_j^2}=\rse{\phi_j\phi_i}{m_i^2}=0,\ i\ne j\ .
\end{equation}
And finally, we have for the non-diagonal entries of the WFRC matrix the renormalization condition
\begin{equation}
\ct{Z_{\phi_i\phi_j}}=\frac{2}{m_{i}^2-m_{j}^2}\se{\phi_i\phi_j}{m_{j}^2}{\text{tad}},\ i\ne j.
    \label{eq:dZ_mixing}
\end{equation}
for each pair of mixing fields. \s


In our case, we thereby have the following mass counterterms for the scalar fields $\phi_i$ ($i=h_1,h_2,A$)
\begin{align}
\ct{m_{\phi_i}^2}&=\se{\phi_i \phi_i}{m_{\phi_i}^2}{\text{tad}} \;.
\end{align}
The WFRCs are given by
\begin{align}
\ctZ{h_ih_i} &=-\pderiv{\se{h_ih_i}{p^2}{\text{tad}}}{p^2}\bigg|_{p^2=m_{h_i}^2}\;\;\;\, , \quad i =1,2 \\
\ctZ{h_ih_j} &=\frac{2}{m_{h_i}^2-m_{h_j}^2}\se{h_ih_j}{m_{h_j}^2}{\text{tad}}\; , \quad i,j=1,2, \; i\ne j \\
\ctZ{A}&=-\pderiv{\se{A}{p^2}{\text{tad}}}{p^2}\bigg|_{p^2=m_{A}^2}\; .
\end{align}
For our one-loop computations we also need to renormalize the electroweak and the fermion  sector. For the massive gauge bosons $(V=W,Z)$ we have the following mass counterterms 
\begin{align}
\ct{m_{V}^2}&=\se{V}{m_{V}^2}{T\,\text{tad}}\; ,
\end{align}
where the superscript 'T' denotes the transverse part of the vector boson self-energy. The WFRCs are given by 
\begin{align}
\ctZ{VV}&=-\pderiv{\se{VV}{p^2}{T\, \text{tad}}}{p^2}\bigg|_{p^2=m_{V}^2}\; , \label{eq:ctZVV}\\
    \ctZ{\gamma\gamma}&=-\pderiv{\se{\gamma\gamma}{p^2}{T}}{p^2}\bigg|_{p^2=0}\; ,\\
    \ctZ{Z \gamma}&=\frac{2}{m_{Z}^2}\se{\gamma Z}{0}{T}\; ,\\
    \ctZ{\gamma Z}&=-\frac{2}{m_{Z}^2}\se{\gamma Z}{m_{Z}^2}{T}\; .
\end{align}
Note, that since the photons do not couple directly to the Higgs bosons, the self-energies involving photons do not receive any tadpole contributions. \s


For the definition of the OS counterterms of the chiral fermions we decompose their  renormalized self-energy into the various contributing chiral components as
\begin{equation}
    \rse{f}{p^2}=\slashed{p}P_L\rseL{f}{p^2}+\slashed{p}P_R\rseR{f}{p^2}+m_f\rseS{f}{p^2} \;,
    \label{eq:renorm_SE_fermion}
\end{equation}
where $\rseL{f}{p^2}$, $\rseR{f}{p^2}$ and $\rseS{f}{p^2}$ represent the renormalized left-chiral, right-chiral, and scalar components, respectively. Applying the OS conditions to \refeq{eq:renorm_SE_fermion} results in the following relations for the fermion mass counterterms $\delta m_f$ and the left- and right-chiral WFRC $\delta Z_f^L$ and $\delta Z_f^R$, respectively, 
\begin{align}
    \ct{m_f}&=\frac{m_f}{2}\left(\se{f}{m_{f}^2}{\text{tad},L}+\se{f}{m_{f}^2}{\text{tad},R}+2\se{f}{m_{f}^2}{\text{tad},S}\right),\\
    \ctZ{f}^R&=-\se{f}{m_{f}^2}{\text{tad},R}-m_f\pderiv{}{p^2}\left(\se{f}{p^2}{\text{tad},L}+\se{f}{p^2}{\text{tad},R}+2\se{f}{p^2}{\text{tad},S}\right)\bigg|_{p^2=m_{f}^2}, \label{IRWFRCR}\\
    \ctZ{f}^L&=-\se{f}{m_{f}^2}{\text{tad},L}
    -m_f\pderiv{}{p^2}\left(\se{f}{p^2}{\text{tad},L}+\se{f}{p^2}{\text{tad},R}+2\se{f}{p^2}{\text{tad},S}\right)\bigg|_{p^2=m_{f}^2} \:.\label{IRWFRCL}
\end{align}
Though not stated explicitly, throughout the whole work we take the real parts of the (renormalized) self-energies and of their derivatives w.r.t.~to $p^2$.

\subsection{Scalar Mixing Angle}
For the renormalization of the mixing angle $\alpha$ we combine the 'KOSY' scheme introduced in \cite{Kanemura:2004mg} with the ATS. In the KOSY scheme the renormalization of the mixing angle is related to the off-diagonal WFRCs. When combined with the ATS, it allows us to define counterterms that are independent of the gauge parameter.\footnote{For a detailed discussion, we refer to \cite{Krause:2016oke}.} In order to unambiguously extract the gauge-parameter independent part of the angular counterterm $\delta \alpha$, we apply the pinch technique \cite{Binosi:2009qm,Cornwall:1989gv,Papavassiliou:1989zd,Papavassiliou:1994pr,Degrassi:1992ue}. With this, the counterterm  in the OS tadpole-pinched scheme \cite{Krause:2016oke} is given by
\begin{align}\label{alpha_os}
\delta \alpha = \frac{1}{2(m_{h_2}^2-m_{h_1}^2)} \left( \left.\left[ \Sigma_{h_1 h_2}^{\text{tad}} (m_{h_2}^2) + \Sigma_{h_2 h_1}^{\text{tad}} (m_{h_1}^2)\right]\right|_{\xi=1} + \Sigma^{\text{add}}_{h_1 h_2} (m_{h_2}^2) + \Sigma^{\text{add}}_{h_2 h_1} (m_{h_1}^2) \right) \;.
\end{align}
The subscript $\xi=1$ indicates that the corresponding self-energies are evaluated in the Feynman gauge. The $\Sigma^{\text{add}}$ is the  additional contribution that arises from the application of the pinch technique. It is explicitly given by
\begin{equation}
    \se{h_ih_j}{p^2}{\text{add}} = -\frac{g^2}{32 \pi^2 c_W^2} O_{h_ih_j} \left(p^2 - \frac{m_{h_i}^2 + 
   m_{h_j}^2}{2}\right) \left(B_0(p^2, m_Z^2, m_Z^2) + 
   2 c_W^2 B_0(p^2, m_W^2, m_W^2)\right)\; ,
\end{equation}
where $g$ is the $U(1)_Y$ coupling constant, $c_W$ is the cosine of the Weinberg angle,  and $B_0$ is the Passarino-Veltman two-point function \cite{Passarino:1978jh}. The  $O_{h_ih_j}$ are defined as
\begin{equation}
    O_{h_ih_j}=\begin{cases}
      \cos^2(\alpha)\; &,\, i=j=1\\
      \sin^2(\alpha)\; &,\, i=j=2\\
      \cos(\alpha) \sin(\alpha); &,\, i\ne j \;.
    \end{cases}
\end{equation}
Alternatively, we can use the $p_\star$ tadpole-pinched scheme \cite{Krause:2016oke}, where all self-energies are evaluated at 
\begin{eqnarray}
p_\star^2 = \frac{m_i^2+m_j^2}{2} \;.
\end{eqnarray}
Here, the additional self-energies $\Sigma^{\text{add}}$ vanish and the angular counterterm takes the simple form
\begin{eqnarray}\label{alpha_pstar}
\delta \alpha = \frac{\left. \Sigma^{\text{tad}}_{h_1 h_2} \left( \frac{m_{h_1}^2+m_{h_2}^2}{2}\right)\right|_{\xi=1}}{m_{h_2}^2-m_{h_1}^2} \;.
\end{eqnarray}
Again, we only take the real parts of the self-energies, wherever they appear. For a detailed derivation of the mixing angle counterterm, we refer to \cite{Krause:2016oke}.


\subsection{Electric Charge}
At the end of subsection \ref{subsec:tadpoles}~we showed that the counterterm $\ct{v}$ for the SM VEV can be written as a function of the counterterms for the $W$ and $Z$ boson masses as well as the counterterm for the electric charge, $\ct{Z_e}$, which we will fix now. \s

The electric charge is defined to be the full electron-positron photon coupling for OS external particles in the Thomson limit, which implies that all corrections to this vertex vanish OS and for zero momentum transfer. The counterterm for the electric charge in terms of the transverse photon-photon and photon-$Z$ self-energies then reads \cite{Denner:1991kt}
\begin{align}
\delta Z_e^{\alpha(0)} = \frac{1}{2} \left.\frac{\partial \Sigma_{\gamma\gamma}^T (p^2)}{\partial p^2}\right|_{p^2 = 0} +\frac{s_W}{c_W} \frac{\Sigma^T_{\gamma Z}(0)}{m_Z^2} \;, \label{eq:alph0}
\end{align}
where $s_W$ denotes the sine of the Weinberg angle. We note that the sign in the second term of Eq.~(\ref{eq:alph0}) differs from the one in \cite{Denner:1991kt} due to our different sign convention in the covariant derivative. Taking the electromagnetic coupling  $\alpha = e^2/(4\pi)$ at zero momentum transfer, $\alpha (0)$, introduces large logarithms due to light fermions $(f\ne t)$. In order to improve the perturbative behaviour, we therefore use the "$G_\mu$ scheme" \cite{Bredenstein:2006rh}, in which 
the electromagnetic coupling constant is derived from the Fermi constant $G_{\mu}$ such that
\begin{equation}   \alpha_{G_\mu}=\frac{\sqrt{2}G_{\mu}m_W^2}{\pi}\left(1-\frac{m_W^2}{m_Z^2}\right)\;.
\end{equation} 
In order to avoid double counting, the corrections that are absorbed in the LO process using $\alpha_{G_\mu}$ must be subtracted from the explicit corrections ${\cal O}(\alpha)$. This is done by subtracting the weak corrections $\Delta r$ to the muon decay \cite{Denner:1991kt,Sirlin:1980nh} from the corrections in the $\alpha (0)$ scheme. We hence redefine the charge renormalization constant as
\begin{align}
\left.\delta Z_e\right|_{G_\mu} = \left.\delta Z_e\right|_{\alpha(0)} - \frac{1}{2} (\Delta r)_{\text{1-loop}} \;,
\end{align}
where $(\Delta r)_{\text{1-loop}}$ is the one-loop expression for $\Delta r$ given by \cite{Denner:1991kt}
\begin{eqnarray}
(\Delta r)_{\text{1-loop}} &=& \left.\frac{\partial \Sigma^T_{\gamma\gamma} (k^2)}{\partial k^2}\right|_{k^2=0} - \frac{c_W^2}{s_W^2} \left( \frac{\Sigma_{ZZ}^T (m_Z^2)}{m_Z^2} - \frac{\Sigma_{WW}^T (m_W^2)}{m_W^2}\right) + \frac{\Sigma_{WW}^T (0)- \Sigma_{WW}^T(m_W^2)}{m_W^2} \nonumber \\
&& -2 \frac{c_W}{s_W} \frac{\Sigma_{\gamma Z}^T (0)}{m_Z^2} + \frac{\alpha}{4\pi s_W^2} \left( 6 + \frac{7-4s_W^2}{2s_W^2} \log c_W^2 \right) \;.
\end{eqnarray}
\s

\subsection{Singlet VEV}
\label{subsec:renorm_vs}
For the renormalization of the singlet VEV $v_S$ we choose a process-dependent renormalization and a derivation thereof, the ZEM scheme, which we introduced in \cite{Azevedo:2021ylf} and used it in the context of the CxSM in \cite{Egle:2022wmq,Egle:2023pbm} to which we also refer for further details. \s

The singlet VEV appears in the trilinear couplings between the visible Higgs bosons $h_i$ ($i=1,2$) and the DM particle $A$,
\begin{align}
\lambda_{h_i AA} = \frac{m_i^2}{v_S} \Bigg\{ \begin{array}{ll} \sin\alpha, & \; i=1 \\
\cos\alpha, & \; i=2 \;.
\label{eq:higgssmcoup}
\end{array}
\end{align}
We therefore can use the partial decay width $\Gamma_{h_i \to AA}$ of the $h_i$ decay into a DM pair  for the renormalization, provided the mass of the decaying particle is above the kinematic threshold,
\begin{align}
m_{h_i} \ge 2 m_A \;. \label{eq:kinem}
\end{align}
We use the following condition to fix the counterterm $\delta v_S$ for the singlet VEV,
\begin{align}
\Gamma^{\text{LO}}_{h_i \to AA} = \Gamma^{\text{NLO}}_{h_i \to AA} \;.
\end{align}
If both $h_i$ have masses above the kinematic threshold, we have two possible renormalization conditions to derive the counterterm $\delta v_S$, which through the OS condition is expressed in terms of other counterterms already introduced in the previous subsections. If we use $\Gamma_{h_1 \to AA}$ ($\Gamma_{h_2 \to AA}$) for the renormalization, $\delta v_S$ will be a function of the mass counterterm $\delta m_{h_1}^2$ ($\delta m_{h_2}^2$), the mixing angle counterterm $\delta \alpha$, the WFRCs $\delta Z_A$, $\delta Z_{h_2 h_1}$ ($\delta Z_{h_1 h_2}$), and $\delta Z_{h_1 h_1}$ ($\delta Z_{h_2 h_2}$). Explicitly, we have
\begin{eqnarray}
\delta v_S^{h_{1}\to AA} &=& v_S \left( - \mbox{Re} \left( \frac{{\cal A}^{\text{VC}}_{h_{1}\to AA}}{\lambda_{h_1AA}} \right) + \frac{\delta m_{h_{1}^2}}{m_{h_{1}}^2} + \cot \alpha \delta \alpha + \delta Z_A \right) \nonumber \\
&& \left. + \frac{\delta Z_{h_1 h_1}}{2} + \frac{\lambda_{h_2AA}}{\lambda_{h_1AA}} \frac{\delta Z_{h_2h_1}}{2} \right.\label{vs_os1} \\
\delta v_S^{h_{2}\to AA} &=& v_S \left( - \mbox{Re} \left( \frac{{\cal A}^{\text{VC}}_{h_2\to AA}}{\lambda_{h_2AA}} \right) + \frac{\delta m_{h_{2}^2}}{m_{h_{2}}^2}  -\tan \alpha \delta \alpha + \delta Z_A \right) \nonumber \\
&& \left. + \frac{\delta Z_{h_2 h_2}}{2} + \frac{\lambda_{h_1AA}}{\lambda_{h_2AA}} \frac{\delta Z_{h_1h_2}}{2} \right.\;.\label{vs_os2}
\end{eqnarray}
Here, the ${\cal A}^{\text{VC}}_{h_i \to AA}$ denote the vertex corrections. \s

In order to avoid the kinematic restriction Eq.~(\ref{eq:kinem}) we also implemented the ZEM scheme for the renormalization of $v_S$. It is inspired by the process-dependent scheme, but with all external momenta set to zero so that the kinematic constraint does not apply. The renormalization condition then reads
\begin{align}
0 = \mbox{Re} \left({\cal A}_{h_i \to AA}^{\text{NLO}} (\{p^2 =0 \}) \right) \;,
\end{align}
where ${\cal A}^{\text{NLO}}_{h_i \to AA}$ denotes the NLO amplitude of the partial decay width $h_i \to AA$ and $p^2=0$ means that all squared external momenta are set to zero. In the ZEM scheme, the counterterm for the singlet VEV for the two possible decay processes is given by \cite{Egle:2022wmq}
\begin{eqnarray}
\delta v_S^{\text{ZEM},h_1 \to AA} &=& v_S \left(- \mbox{Re} \left( \frac{{\cal A}^{\text{VC}}_{h_1 \to AA} (\{ p^2 = 0\}) + {\cal A}^{\text{Leg}}_{h_1 \to AA} (\{ p^2 = 0\})}{\lambda_{h_1AA}}\right) \right. \nonumber \\
&+& \left. \cot\alpha \delta \alpha - \delta Z_A - \frac{2 \delta m_A^2}{m_A^2} - \frac{\delta Z_{h_1 h_1}^{\text{pinched}}}{2} - \tan\alpha \frac{\delta Z_{h_1 h_2}^{\text{pinched}}}{2} \right) \label{vs_zem1}\\
\delta v_S^{\text{ZEM},h_2 \to AA} &=& v_S \left(- \mbox{Re} \left( \frac{{\cal A}^{\text{VC}}_{h_2 \to AA} (\{ p^2 = 0\}) + {\cal A}^{\text{Leg}}_{h_2 \to AA} (\{ p^2 = 0\})}{\lambda_{h_1AA}}\right) \right. \nonumber \\
&-& \left. \tan\alpha \delta \alpha - \delta Z_A - \frac{2 \delta m_A^2}{m_A^2} - \frac{\delta Z_{h_2 h_2}^{\text{pinched}}}{2} - \tan\alpha \frac{\delta Z_{h_2 h_1}^{\text{pinched}}}{2} \right)\label{vs_zem2} \;.
\end{eqnarray}
We note that now the NLO leg corrections ${\cal A}^{\text{Leg}}$ appear as we set $p^2=0$, which implies that the NLO leg corrections are not canceled by the corresponding counterterms that are defined through the OS scheme. We furthermore use the pinched WFRCs $\delta Z_{h_i h_j}^{\text{pinched}}$ ($i,j=1,2$) in order to get rid of the gauge parameter dependence that is otherwise introduced through the ZEM scheme.


\section{Treatment of the Infrared Divergences}
\label{sec:IRDivergencies}

\begin{figure}[!ht]

\scalebox{0.6}{



\hspace{6.4cm}
\begin{tikzpicture}
\begin{feynman}
\vertex at (-1.4,1)(a){\(\text{DM}\)};
 \vertex[blob] (b) at ( 0, 0) {} ;
\vertex at (1.4,0) (d);
\vertex at (-1.4,-1) (c){\(\text{DM}\)};
\vertex at (2.8,-1) (w){\(W^{-},Z,\bar{f},h_{k}\)};
\vertex at (2.35,0.7) (w1);
\vertex at (2.8,1) (u) {\(W^{+},Z,f,h_{j}\)};

\diagram* {
(a) -- [scalar] (b),
(c) -- [scalar] (b),
(b) -- [scalar, edge label'=\(h_{i}\)] (d),
(d) -- [] (w),
(d) -- [] (u),

};
\end{feynman}
\end{tikzpicture}
\hspace{0.5cm}

\begin{tikzpicture}

\begin{feynman}
\vertex at (-1.4,1)(a){\(\text{DM}\)};
 \vertex  at ( 0, 0) (b) ;
\vertex (d) at (1.4,0) ;
\vertex[blob] (m) at (0.7,0) {};
\vertex at (-1.4,-1) (c){\(\text{DM}\)};
\vertex at (2.8,-1) (w){\(W^{-},Z,\bar{f},h_{k}\)};
\vertex at (2.35,0.7) (w1);
\vertex at (2.8,1) (u) {\(W^{+},Z,f,h_{l}\)};

\diagram* {
(a) -- [scalar] (b),
(c) -- [scalar] (b),
(b) -- [scalar, edge label'=\(h_{i}\)] (m) -- [scalar, edge label=\(h_{j}\)] (d),
(d) -- [] (w),
(d) -- [] (u),

};
\end{feynman}
\end{tikzpicture}

}
\scalebox{0.6}{

\begin{tikzpicture}
\begin{feynman}
\vertex at (-1.4,1)(a){\(\text{DM}\)};
 \vertex  at ( 0, 0) (b) ;
\vertex[blob] (d) at (1.4,0) {} ;
\vertex at (-1.4,-1) (c){\(\text{DM}\)};
\vertex at (2.8,-1) (w){\(W^{-},Z,\bar{f},h_{k}\)};
\vertex at (2.35,0.7) (w1);
\vertex at (2.8,1) (u) {\(W^{+},Z,f,h_{j}\)};

\diagram* {
(a) -- [scalar] (b),
(c) -- [scalar] (b),
(b) -- [scalar, edge label'=\(h_{i}\)] (d),
(d) -- [] (w),
(d) -- [] (u),

};
\end{feynman}
\end{tikzpicture}

}

\vspace*{0.5cm}

\scalebox{0.6}{
\hspace*{7.3cm}

\begin{tikzpicture}
\begin{feynman}
\vertex at (-1.4,1)(a) {\(\text{DM}\)};
\vertex[blob](b) at (0.7,0) {};
\vertex at (-1.4,-1) (c) {\(\text{DM}\)};
\vertex at (2.8,-1) (w) {\(h_{j}\)};
\vertex at (2.8,1) (u) {\(h_{i}\)};

\diagram* {
(a) -- [scalar] (b),
(c) -- [scalar] (b),
(b) -- [scalar] (w),
(b) -- [scalar] (u),

};
\end{feynman}
\end{tikzpicture}

\hspace*{0.5cm}







\begin{tikzpicture}
\begin{feynman}
\vertex at (-1.4,1)(a) {\(\text{DM}\)};
\vertex at (0.7,1)  (b);
\vertex at (0.7,-1) (k);
\vertex[blob] (m) at (0.7,0) {};
\vertex at (-1.4,-1) (c)  {\(\text{DM}\)};
\vertex at (2.8,-1) (w) {\(h_{j}\)};
\vertex at (2.8,1) (u) {\(h_{i}\)};

\diagram* {
(a) -- [scalar] (b),
(c) -- [scalar] (k),
(k) -- [scalar] (w),
(b) -- [scalar] (u),
(b) -- [scalar, edge label'=\(\text{DM}\)]  (m) --[scalar, edge label'=\(\text{DM}\)] (k),

};
\end{feynman}
\end{tikzpicture}

\hspace*{0.5cm}

\begin{tikzpicture}
\begin{feynman}
\vertex at (-1.4,1)(a) {\(\text{DM}\)};
\vertex at (0.7,1)  (b);
\vertex[blob] (k) at (0.7,-1) {};
\vertex at (-1.4,-1) (c)  {\(\text{DM}\)};
\vertex at (2.8,-1) (w) {\(h_{j}\)};
\vertex at (2.8,1) (u) {\(h_{i}\)};

\diagram* {
(a) -- [scalar] (b),
(c) -- [scalar] (k),
(k) -- [scalar] (w),
(b) -- [scalar] (u),
(b) -- [scalar, edge label'=\(\text{DM}\)]  (k),

};
\end{feynman}
\end{tikzpicture}

}
\vspace*{0.5cm}

\scalebox{0.6}{
\hspace{6.9cm}




\begin{tikzpicture}
\begin{feynman}
\vertex at (-1.4,1)(a) {\(\text{DM}\)};
\vertex[blob] (b) at (0.7,1) {};
\vertex at(0.7,-1) (k);
\vertex at (-1.4,-1) (c)  {\(\text{DM}\)};
\vertex at (2.8,-1) (w) {\(h_{j}\)};
\vertex at (2.8,1) (u) {\(h_{i}\)};

\diagram* {
(a) -- [scalar] (b),
(c) -- [scalar] (k),
(k) -- [scalar] (w),
(b) -- [scalar] (u),
(b) -- [scalar, edge label'=\(\text{DM}\)]  (k),

};
\end{feynman}
\end{tikzpicture}
\hspace{0.5cm}
\begin{tikzpicture}
\begin{feynman}
\vertex at (-1.4,1)(a){\(\text{DM}\)};
\vertex at(0,0) (b);
\vertex at (1.4,0) (d);
\vertex at (2.1,0.395) (r);
\vertex at (2.8,0.2) (t);
\vertex at (3.2,0.2) {\(g,\gamma\)};
\vertex at (-1.4,-1) (c) {\(\text{DM}\)};
\vertex at (2.8,-1) (w);
\vertex at (3.2,-1) {\(W,\bar{f}\)};
\vertex at (2.35,0.7) (w1);
\vertex at (2.8,1) (u); 
\vertex at (3.2,1) (u) {\(W,f\)};

\diagram* {
(a) -- [scalar] (b),
(c) -- [scalar] (b),
(b) -- [scalar, edge label'=\(h_{i}\)] (d),
(d) -- [] (w),
(d) -- [] (u),
(r) -- [boson]  (t),

};
\end{feynman}
\end{tikzpicture}
 \hspace{0.5cm}
\begin{tikzpicture}
\begin{feynman}
\vertex at (-1.4,1)(a){\(\text{DM}\)};
\vertex at(0,0) (b);
\vertex at (1.4,0) (d);
\vertex at (2.1,-0.5) (r);
\vertex at (2.8,-0.2) (t);
\vertex at (3.2,-0.2) {\(g,\gamma\)};
\vertex at (-1.4,-1) (c) {\(\text{DM}\)};
\vertex at (2.8,-1) (w);
\vertex at (3.2,-1) {\(W,\bar{f}\)};
\vertex at (2.35,0.7) (w1);
\vertex at (2.8,1) (u); 
\vertex at (3.2,1) (u) {\(W,f\)};

\diagram* {
(a) -- [scalar] (b),
(c) -- [scalar] (b),
(b) -- [scalar, edge label'=\(h_{i}\)] (d),
(d) -- [] (w),
(d) -- [] (u),
(r) -- [boson]  (t),

};
\end{feynman}
\end{tikzpicture}
}
\caption{Generic diagrams contributing at NLO to DM annihilation into a massive gauge boson pair $V$ ($V=W^\pm, Z$), a fermion pair $f$, or a Higgs boson pair $h_{i,j}$ ($i,j=1,2$). 
The shaded blobs denote all  possible loop and counterterm insertions in the CxSM. The latter two diagrams with the real radiation of a gluon $g$ or photon $\gamma$, respectively, have to be included to render the NLO result IR finite. \label{fig:sample}}
\end{figure}
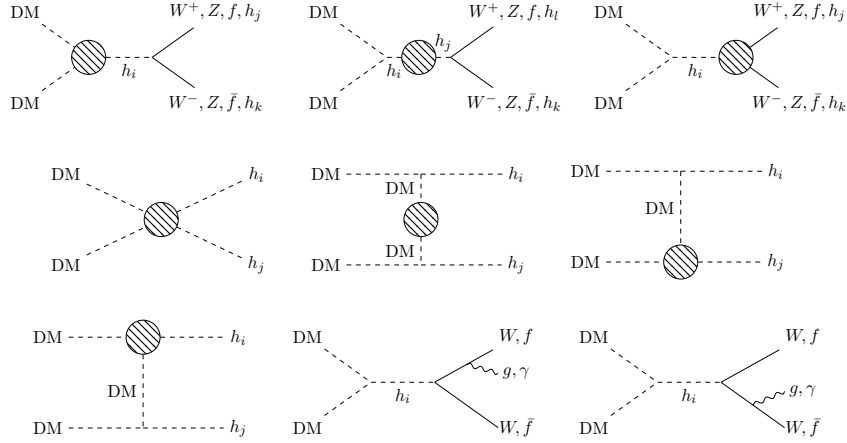

Besides UV divergences we can also encounter infrared (IR) divergences. These arise from loop diagrams with massless particles in the loop. According to the Kinoshita-Lee-Nauemberg theorem IR safety is guaranteed after including the real corrections   \cite{Kinoshita:1962ur, Lee:1964is}. The total UV- and IR-convergent cross section then reads
\begin{equation}
\sigma_{\text{total}}=\underbrace{\int\vert \mathcal{M}_{\text{LO}}\vert^{2}\dd\Pi_{2}}_{\sigma_{\text{LO}}}+\underbrace{\int 2\text{Re} (\mathcal{M}_{\text{LO}}\mathcal{M}^{*}_{1\text{loop}})\dd\Pi_{2}}_{\sigma_{\text{virtual}}}+\underbrace{\int{\vert \mathcal{M}_{\text{real}}}\vert^{2}\dd\Pi_{3}}_{\sigma_{\text{real}}}\,.
\end{equation}
 Here, $\sigma_{\text{LO}}$, $\sigma_{\text{virtual}}$ and $\sigma_{\text{real}}$ denote the LO, the virtual, and the real cross section, respectively, and $\dd\Pi_{2}$ and $\dd\Pi_{3}$ are the Lorentz-invariant phase space measures of the two- and three-particle phase space, respectively. 
While  $\sigma_{\text{total}}$ is IR finite, the virtual cross section $\sigma_{\text{virtual}}$ and the real cross section $\sigma_{\text{real}}$ are separately IR divergent. The former divergence arises from the presence of massless gauge bosons (photons or gluons) in the loops, in particular from IR-divergent derivatives of the self-energies that enter the wavefunction renormalization constants Eqs.\,(\ref{eq:ctZVV}), (\ref{IRWFRCR}), and (\ref{IRWFRCL}), as well as from triangle diagrams. The IR-divergences of the real corrections stem from the divergent phase space of the emitted massless gauge boson in the soft limit. We show sample diagrams for the gauge boson final states in Fig.~\ref{fig:sample}. \s

To achieve a Lorentz-invariant cancellation of the IR divergences between the real and virtual corrections, dimensional regularization should be applied, since the integration has to be performed in $D=4-2\epsilon$ dimensions. This prevents a straightforward numerical integration. To circumvent this problem, the Catani-Seymour dipole subtraction formalism was developed \cite{Catani_1997}. Here an auxiliary counterterm $\dd\sigma_{\text{A}}$ is introduced, which mimics the divergent behavior of the real corrections. In order to achieve separate IR finiteness in the real and virtual corrections, the counterterm is subtracted from the real corrections while it is added to the virtual corrections after integrating over the one-particle phase space of the particle radiated in the real process,
 \begin{equation}
\sigma_{\text{NLO}}=\int \biggl{[}\bigl{(}\dd\sigma_{\text{real}}\bigr{)}_{\epsilon=0}-\bigl{(}\dd\sigma_{\text{A}}\bigr{)}_{\epsilon=0}\biggr{]}+\int\biggl{[}\dd\sigma_{\text{virtual}}+\int\dd\sigma_{\text{A}}\biggr{]}_{\epsilon=0}\,.
\label{Catani}
\end{equation} 
The integrals successively denote the integration over the 3-particle, 2-particle and 1-particle phase space. The subscript $\epsilon=0$ indicates that the limit to four dimensions is taken before the integration over the 3-particle and the 2-particle phase space, respectively, is performed. \s

 The auxiliary counterterm can be constructed in such a way that it factorizes in the LO amplitude $\dd\sigma_{\text{LO}}$ and the dipoles $\dd V_{\text{dipoles}}$. The latter depend on the process of interest, namely on the real corrections and how the real radiation is realized. The dipoles as well as the integrated dipoles are given in Refs.~\cite{Catani_2002, Sch_nherr_2018} for every possible scenario. With these,  Eq.~(\ref{Catani}) can then be evaluated numerically in four dimensions since  the 2- and 3-particle phase spaces are now IR convergent by themselves. 

\FloatBarrier


\section{Program Description \label{sec:progdescr}}
The new calculation of the relic density at NLO in the strong and electroweak interactions has been implemented in the existing code \texttt{RelExt} \cite{Capucha:2025iml}. The code can be downloaded from 
\begin{center}
\url{https://github.com/jplotnikov99/RelExt/tree/NLO}
\end{center}
Also the manual Ref.~\cite{Capucha:2025iml} can be found there, which contains details on the installation and the usage of the code. 
We have extended the settings in the file \texttt{main.cpp} to include the possiblity of the computation of the NLO QCD and EW corrections to the relic density. In order to compute the involved Passerino-Veltmann functions \cite{Passarino:1978jh}, we use the public library \texttt{COLLIER} \cite{Denner_2017}. To evaluate the three-body phase space integral for the computation of the real emission cross section, we use the public library \texttt{Cuba}\cite{Hahn_2005}. During the initial setup of \texttt{RelExt}, the libraries \texttt{Cuba-4.2.2} and \texttt{COLLIER-1.2.8} are downloaded and linked automatically. To install \texttt{RelExt}, go to the folder, where \texttt{RelExt} is located and use the following commands
\begin{lstlisting}
mkdir build && cd build
cmake .. 
make
\end{lstlisting}
Note that the current version of the NLO extension supports only the CxSM. To use the other models at LO, the master branch of \texttt{RelExt} must be used. \s

Examples for input and \texttt{main.cpp} files together with the corresponding output files are given on the webpage. 
For the convenience of the user and to present the various implemented options in the computation of the NLO corrections, we present in the Appendix \ref{app:bp} for a point obtained from our scan described in the next section, a sample \texttt{main.cpp} file together with a presentation of its output. 
\section{Parameter Scan \label{sec:param}}
For the subsequent discussion of the impact of the included higher-order corrections on the relic density we generated a sample of parameter points that fulfill all relevant theoretical and experimental constraints. They are included in our code \texttt{ScannerS} \cite{Coimbra:2013qq,Muhlleitner:2020wwk} that we use for the scan. Here, we briefly describe the applied constraints and refer the reader for more details to previous publications \cite{Egle:2022wmq,Egle:2023pbm}. \s

On the theory side we include the constraints from boundedness from below, the stability of the vacuum and perturbative unitarity. For the latter to be ensured, we require the eigenvalues of the scattering matrix of all possible two-to-two scalar scatterings to be below $8\pi$ \cite{Lee:1977eg}.  \s

As for the experimental constraints, the $\rho$ parameter is equal to one at tree level and there are no flavour-changing neutral couplings in our model. For the compatibility with the EW precison data we require the $S,T,U$ parameters \cite{Peskin:1991sw} to be consistent with the measured quantities at 95\% C.L. Compatibility of the SM-like Higgs boson with the LHC Higgs data and of the non-SM-like one with the exclusion bounds is checked through the link to \texttt{HiggsTools} \cite{Bahl:2022igd}. We require that the rates of the SM-like Higgs are in agreement with the experimental data at the $2\sigma$ level. 
At this stage, we do not yet apply constraints on the DM relic density. This allows us to investigate if the computed higher-order corrections to the relic density may allow points previously excluded at LO, and vice versa. As for DM direct detection, we note that in this model the LO cross section is proportional to the velocity of the DM particle and therefore negligible \cite{Gross:2017dan,Azevedo:2018exj}. The NLO contribution may play a role in future direct detection constraints \cite{Glaus:2020ihj}. \s

\begin{table}[htbp]
\centering
\small
\begin{tabular}{ccccccc}
\toprule
$m_W$ & $m_Z$ & $m_{h}$ & $\alpha_\text{EW}(m_Z)$ & $\alpha_{\text{s}}(m_Z)$ & $m_b^{\text{pole}}$ & $m_t^{\text{pole}}$ \\
\midrule
 79.8244 &  91.1876 & 125.09 & 1/127.9 & 0.1184  & 4.7 & 172 \\
\bottomrule
\end{tabular}
\caption{SM input parameters: The $W$ and $Z$ boson mass, the mass of the SM-like Higgs boson, the electromagnetic coupling at the $Z$ mass scale, the strong coupling constant at the $Z$ mass scale, and the bottom and top quark pole masses. All masses are given in GeV.}
\label{tab:SMparams}
\end{table}
The chosen values for the SM parameters are given in Tab.~\ref{tab:SMparams}, and the parameter ranges of our scan values are  summarized in Tab.~\ref{tab:rangesofscan}. We denote the mass of the non-SM-like Higgs by $m_S$. It need not necessarily be the lightest Higgs $h_1$ as in principle we can have scenarios where the SM-like Higgs, denoted by $h$ in the following, can be the lighter or the heavier of the two visible scalars. All results that we show in the following, however, only include scenarios where $m_S > m_h$.
We start our scan at a dark matter mass of $m_A=62.5$~GeV so that no SM-Higgs decays in dark matter particles are kinematically possible.
\begin{table}[h!]
\centering
\begin{tabular}{ccc}
\toprule
Parameter & \multicolumn{2}{c}{Range} \\
\cmidrule{2-3}
 & {Lower} & {Upper} \\
\midrule
$m_S$ & 126 GeV & 1000 GeV\\ 
$m_A$ & 62.5 GeV & 1000 GeV \\
$v_S$ & 1 GeV & 1000 GeV\\
$\alpha$ & $-$1.57 &  1.57 \\
\bottomrule
\end{tabular}
\caption{Scan ranges of the parameters used in the \texttt{ScannerS} scan for the generation of the parameter points used in the numerical analysis. The mass of the non-125~GeV Higgs boson is denoted by $m_S$. 
\label{tab:rangesofscan}}
\end{table}


\section{Numerical Analysis}
\label{sec:numericalanalysis}
In the following numerical analysis we present the size and impact of the NLO QCD and EW corrections on the relic density. We furthermore discuss the remaining theoretical uncertainties due to missing higher-order corrections.

\subsection{Relative Corrections to the Relic Density}
Before showing our results on the relative corrections to the relic density, we want to make a few general considerations about the expected corrections. 
Denoting by $\Omega_{\text{LO}}$ the LO relic density and by $\Omega_{\text{NLO}}$ the pure NLO correction to the relic density, i.e.~the sum of the  virtual corrections and the counterterms plus the real corrections, where necessary,  we define the total corrected relic density at NLO as
\begin{eqnarray}
\Omega_{\text{corr}} = \Omega_{\text{LO}} + \Omega_{\text{NLO}} = \Omega_{\text{LO}} \left( 1 + \frac{\Omega_{\text{NLO}}}{\Omega_{\text{LO}}} \right) \equiv \Omega_{\text{LO}} \left( 1 + \kappa_{\Omega} \right) \;,
\end{eqnarray}
where we introduced the relative correction to the relic density,
\begin{eqnarray}
\kappa_\Omega \equiv \frac{\Omega_{\text{NLO}}}{\Omega_{\text{LO}}} \;.
\end{eqnarray}
The relic density is approximately proportional to the inverse of the thermally averaged cross section defined in Eq.~(\ref{eq:TAC}). Here, for convenience of the notation, we denote the TAC simply by $\sigma$, hence
\begin{eqnarray}
\Omega \sim \frac{1}{\sigma} \;.
\end{eqnarray}
The LO and the NLO values of the TAC are derived from Eq.~(\ref{eq:TAC}) by replacing the DM annihilation cross section $\sigma_{ij} (s)$ with the LO value $\sigma^{\text{LO}}_{ii}$ defined in Eq.~(\ref{eq:siglo}), and the NLO value $\sigma^{\text{NLO}}_{ii}$ defined in Eq.~(\ref{eq:sigmacorr}). 
With this, we can then derive the relation between the relative correction $\kappa_\Omega$ to the relic density and the relative correction to the TAC,
\begin{eqnarray}
\kappa_\sigma \equiv \frac{\sigma ^{\text{NLO}}}{\sigma^{\text{LO}}}
\end{eqnarray}
as
\begin{eqnarray}
\kappa_{\Omega} = \frac{\Omega_{\text{NLO}}}{\Omega_{\text{LO}}} = \frac{\Omega_{\text{corr}}-\Omega_{\text{LO}}}{\Omega_{\text{LO}}} \sim \frac{\frac{1}{\sigma^{\text{LO}}+\sigma^{\text{NLO}}}-\frac{1}{\sigma^{\text{LO}}}}{\frac{1}{\sigma^{\text{LO}}}} 
= \frac{1}{1+\frac{\sigma^{\text{NLO}}}{\sigma^{\text{LO}}}} -1 \equiv \frac{1}{1+\kappa_\sigma} -1 \;.
\label{eq:kapomega}
\end{eqnarray}
Hence,
\begin{eqnarray}
\kappa_\Omega \sim \frac{1}{1+\kappa_\sigma} -1 \;.
\end{eqnarray}
More specifically, this means that for negative relative corrections of the thermally averaged cross section the relative correction to the relic density grows very quickly. We have 
\begin{eqnarray}
\kappa_\sigma \in \{0,...,-1 \} \quad \leadsto \quad \kappa_\Omega \in \{0,...,+\infty\} \;.
\end{eqnarray}
Positive relative corrections of the thermally averaged cross section, on the other hand, induce a much more moderate decrease in the relative correction to the relic density, which cannot fall below -100\%, i.e.
\begin{eqnarray}
\kappa_\sigma \in \{0,...,+\infty \} \quad \leadsto \quad \kappa_\Omega \in \{0,...,-1\} \;.
\end{eqnarray}
Taking as an example relative corrections to the relic density between $-40\%$ and $+40\%$, this roughly corresponds to relative corrections of the thermally averaged cross section between $+67\%$ and $-29\%$.

\subsection{QCD Corrections to the Relic Density}
We start by discussing the QCD corrections, which also allow us to do a simple cross-check of our results with the existing literature.
\begin{figure}[h!]\centering\includegraphics[width=0.49\textwidth]{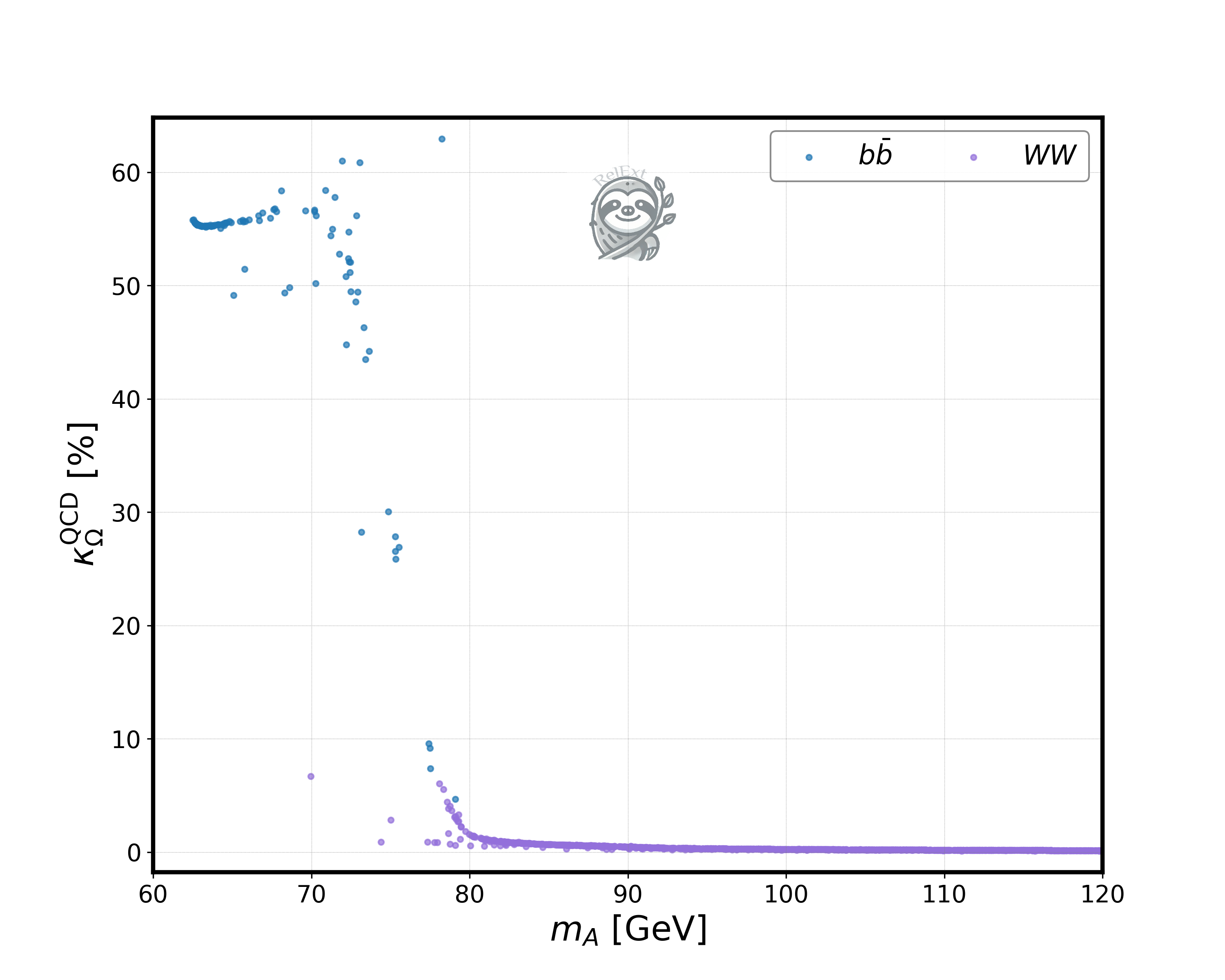}
\hspace*{-0.5cm}\includegraphics[width=0.49\textwidth]{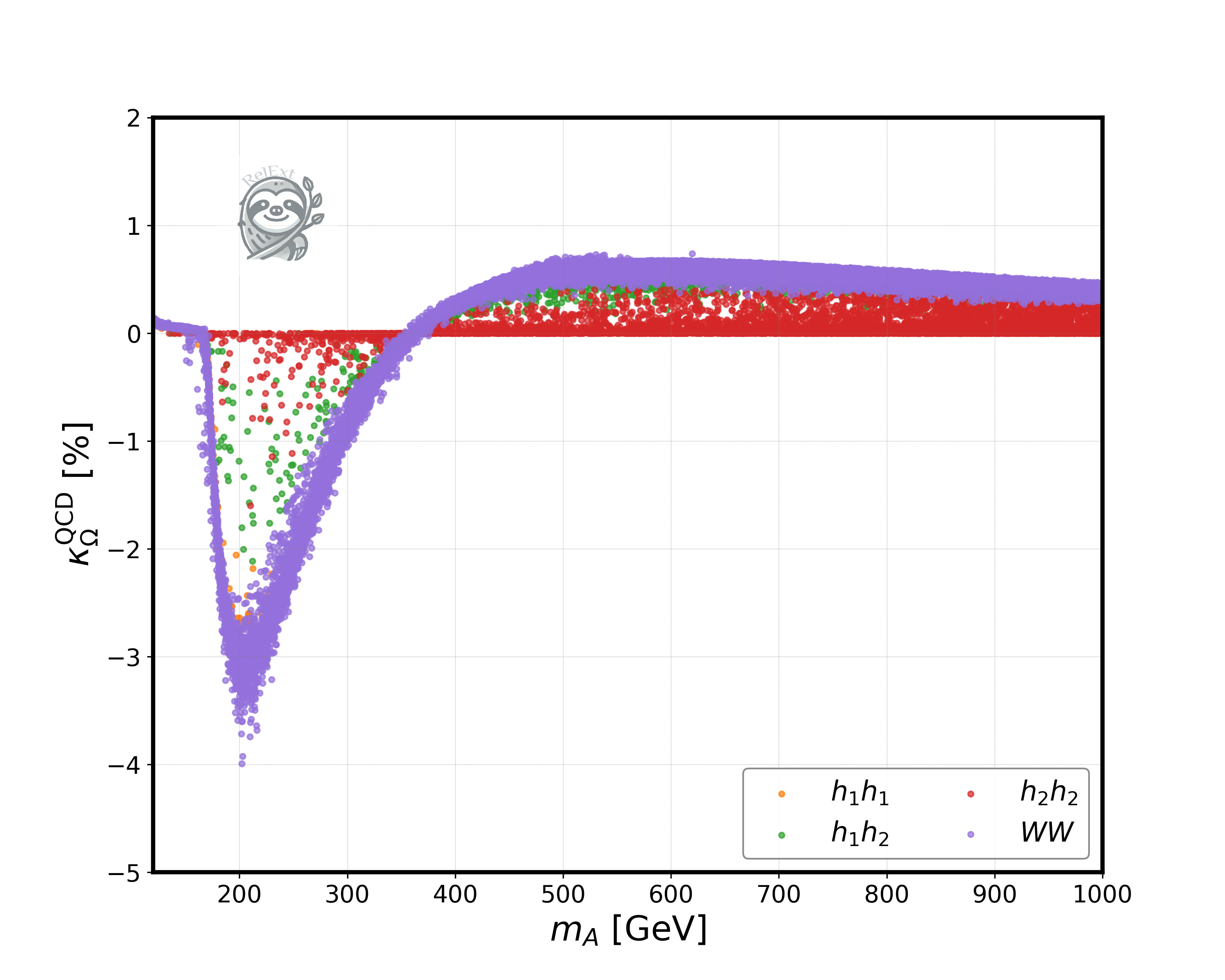}
    \caption{ Relative QCD corrections to the relic density, $\kappa_\Omega^\text{QCD}$, as a function of the DM mass $m_A$ for the mass ranges $62\;\mathrm{GeV} \leq m_A \leq 120\;\mathrm{GeV}$ (left) and $120\;\mathrm{GeV} \leq m_A \leq 1\;\mathrm{TeV}$ (right). The color code indicates the dominant channel in the annihilation cross section for the individual parameter point.}
\label{fig:relicqcd}
\end{figure}
In Fig.~\ref{fig:relicqcd}, we show the relative QCD  corrections $\kappa_\Omega^{\text{QCD}}$ to the relic density as a function of the DM mass $m_A$. On the left panel, we show the mass range $60\;\mathrm{GeV}\leq m_A \leq 120\;\mathrm{GeV}$, while on the right panel, the mass range $120\;\mathrm{GeV}\leq m_A\;\leq 1\;\mathrm{TeV}$ is plotted.  The color code indicates the dominant channels in the annihilation cross section for the individual parameter point.  The channels are dark matter annihilation into  $b\bar{b}$ (blue), $W^+W^-$ (purple), $h_1h_1$ (orange), $h_1h_2$ (green), and $h_2h_2$ (red) pairs. The dominance of the channels can be understood from the kinematics. For DM masses up to about 78~GeV, the dominant channel is annihilation into $b\bar{b}$ since the $WW$ channel is still closed and the maximum coupling is given by the heaviest kinematically accessible fermion, which has the largest Yukawa coupling. Once the $WW$ channel is kinematically accessible, it takes the lead. Above the thresholds for scalar pair production ($m_A >$~125~GeV), 
also Higgs pair final states play a role, with the importance varying from parameter point to parameter point. Here, not only trilinear Higgs couplings, but also quartic couplings between two DM and two Higgs particles play a role. These Higgs trilinear and quartic couplings have an involved dependence on the model parameters. \s

In the region where $b\bar{b}$ final states dominate, we observe relative corrections of up to $\sim 65\%$, whereas otherwise the corrections are rather small and vary between about -4\% and $+0.8$\%. The large  corrections can be understood as follows. As the dominant contribution is given by the $b\bar{b}$ channel, the TAC is dominated by the following term (cf. Eq.~\eqref{eq:TAC}), 
\begin{equation}
    \langle \sigma v\rangle \propto \int\sqrt{s}p^2_{AA}\sigma^{\text{NLO QCD}}_{AA\to b\bar{b}}\, K_1\left(\frac{\sqrt{s}x}{m_A}\right)\mathrm{d}s\;.
\end{equation}
Here, $\sigma_{AA\to b\bar{b}}^{\text{NLO QCD}}$ denotes cross section of DM annihilation into $b\bar{b}$ at NLO QCD.  As neither the DM particle nor the Higgs mediators carry a colour charge, the QCD corrections only appear in the $h_i b\bar{b}$ vertices ($i=1,2$) and the real corrections, cf.~Fig.~\ref{fig:sample}. The QCD corrections to DM annihilation into $b\bar{b}$ can hence be approximated by the QCD corrections to the Higgs boson decays into $b\bar{b}$. Since the Higgs constraints push valid parameter points into the SM limit the contributions from the non-SM-like Higgs are suppressed. Moreover, we consider scenarios where the SM-like Higgs is the lighter of the the two Higgs bosons, so that the $s$-channel contribution of the heavier non-SM-like Higgs is further suppressed. We can hence focus on the corrections to the SM-like Higgs bosons $h$. The QCD corrections to the SM-like Higgs decay into a $b$-quark pair are given by,
cf.~\cite{Kniehl:1994ju,Kwiatkowski:1994cu,Spira:1995gc},
\begin{eqnarray}
\Gamma^{\text{NLO QCD}}_{h\to b\bar{b}} = \Gamma^{\text{LO}}_{h\to b\bar{b}} (1 + \kappa^{\text{QCD}}) \;,
\end{eqnarray}
with
\begin{eqnarray}
\kappa^{\text{QCD}} = C_F \frac{\alpha_s}{\pi} \left( \frac{9}{4} - \frac{3}{2} \ln \left( \frac{m_{h_{\text{SM}}}^2}{m_b^2} \right) \right) \approx -38\% \,,
\end{eqnarray}
where we used $C_F = 4/3$, the bottom pole mass $m_b=4.7$~GeV, and $\alpha_s=0.118$. We note that in \texttt{RelExt}, we actually use the running $\overline{\mbox{MS}}$ bottom quark mass, which barely changes this result. Inserting this as value for $\kappa_\sigma$ into Eq.~(\ref{eq:kapomega}), we obtain
\begin{eqnarray}
\kappa_\Omega^{\text{QCD}} \approx 61\%. 
\end{eqnarray}
This estimate is in agreement with the largest corrections that we found. With the increasing importance of the other annihilation channels that do not receive QCD corrections, the QCD corrections on the relic density decrease.

\subsection{Impact of the Electroweak and QCD Corrections}
\begin{figure}[h!]\centering\includegraphics[width=0.49\textwidth]{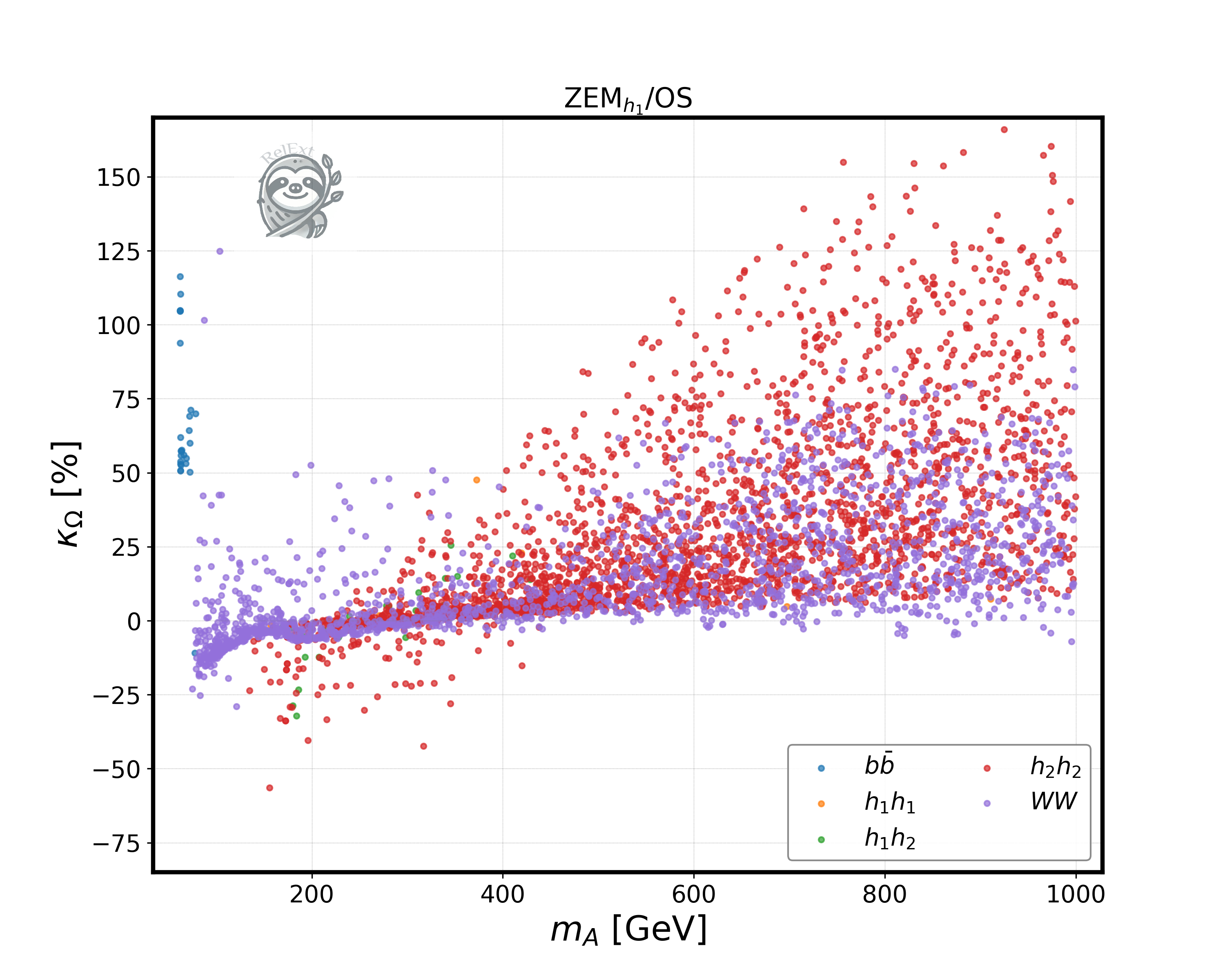}\hfill\includegraphics[width=0.49\textwidth]{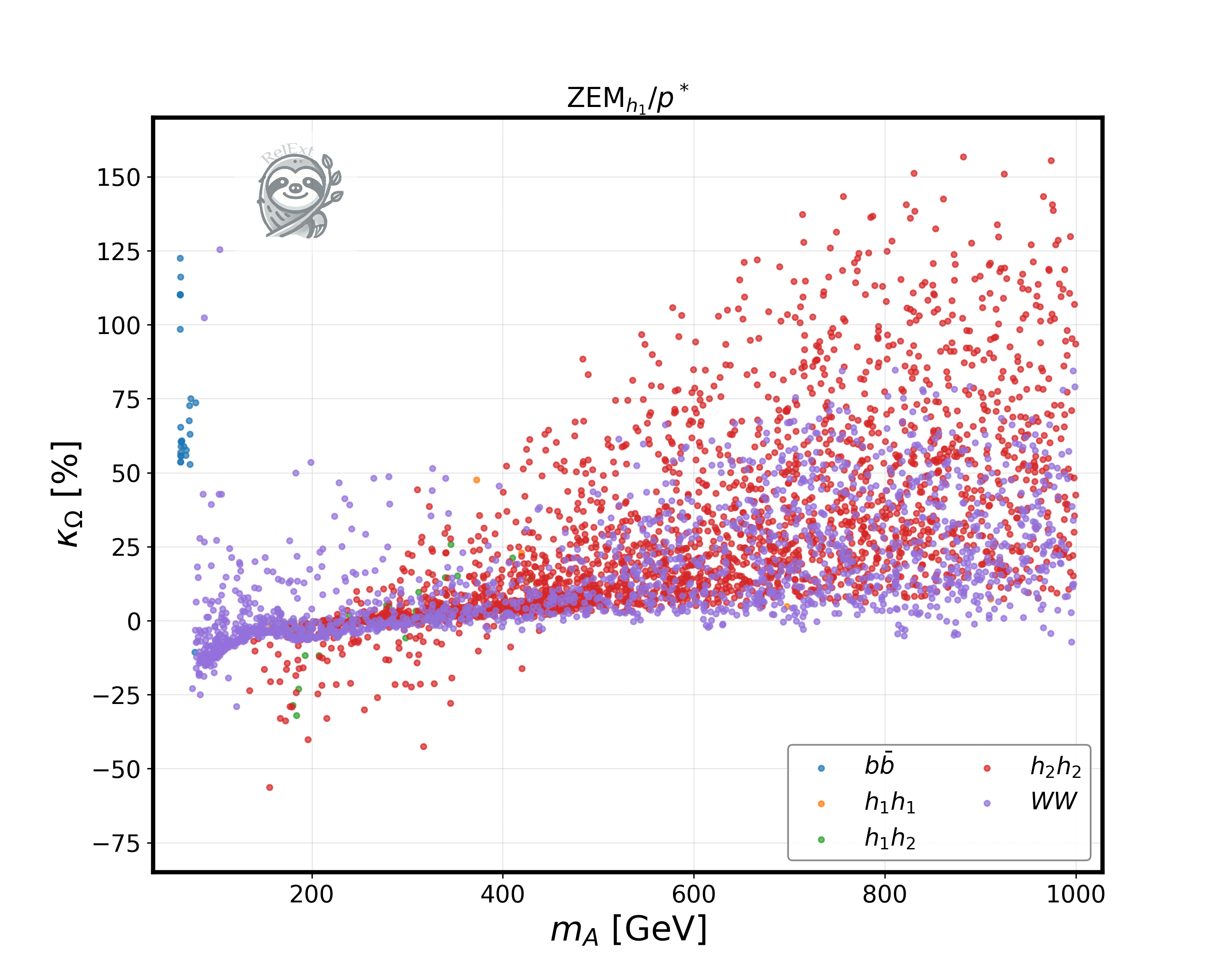}\\[-1ex]\includegraphics[width=0.49\textwidth]{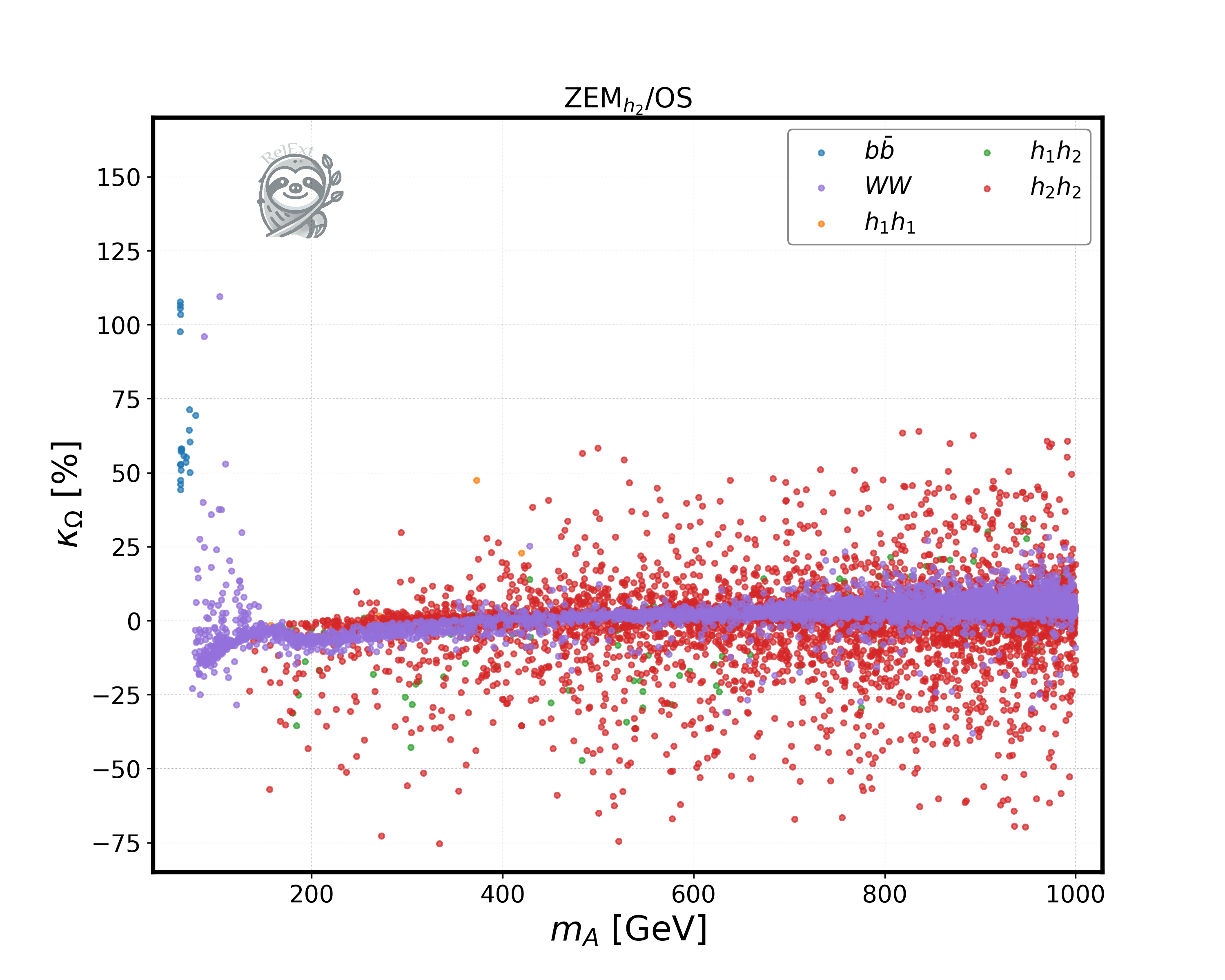}\hfill\includegraphics[width=0.49\textwidth]{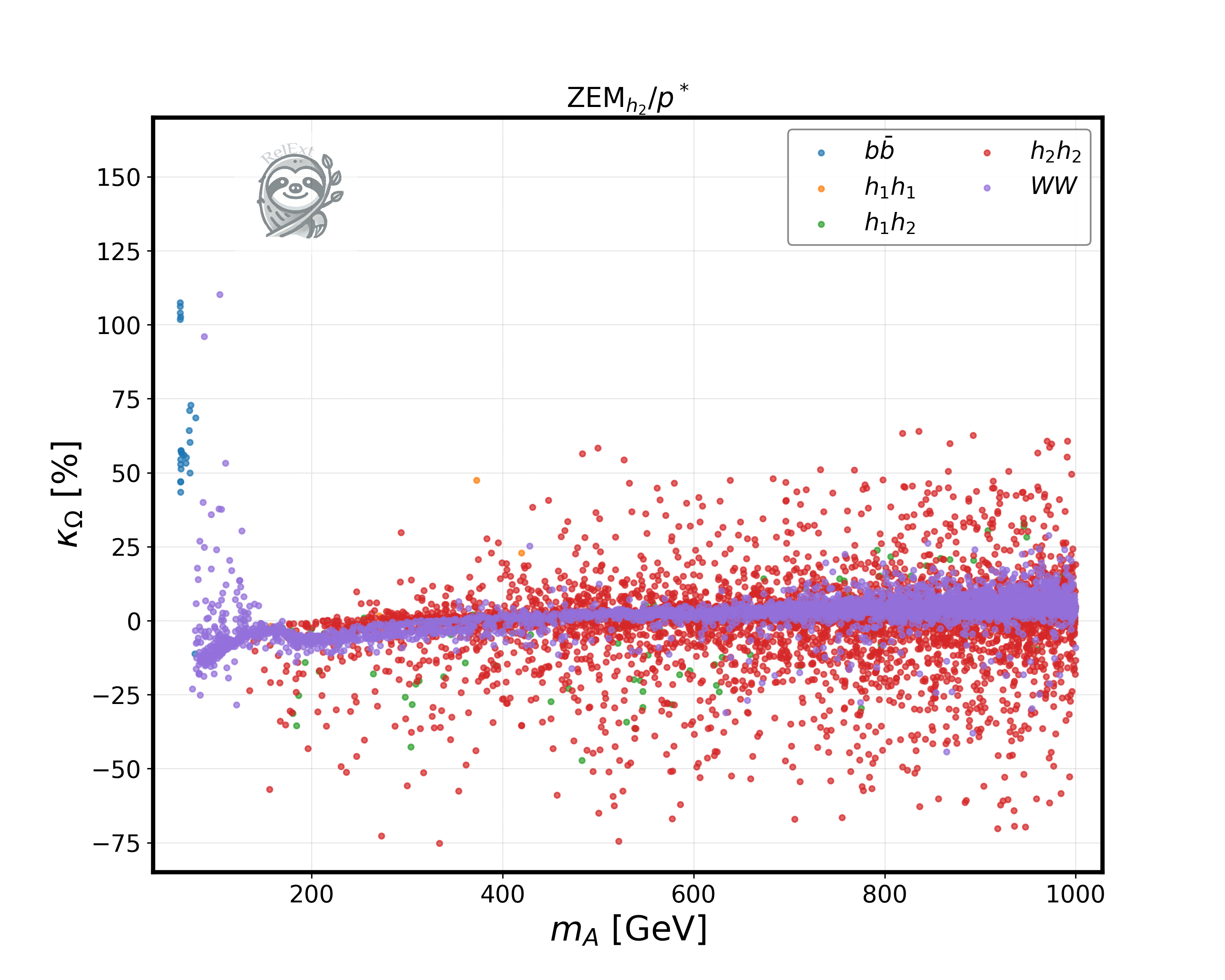}
\caption{Relative QCD plus EW corrections $\kappa_\Omega$ to the relic density as a function of the DM mass $m_A$. The color code indicates the dominant annihilation channel, given by $b\bar{b}$ (blue), $W^+W^-$ (purple), $h_2h_2$ (red), $h_1h_1$ (orange), and $h_1h_2$ (green). 
Upper: The chosen renormalization scheme on the left (right) for $\delta v_s$ is the $\mathrm{ZEM}_{h_1}$ scheme and for $\delta \alpha$ the OS tadpole-pinched scheme ($p^*$ tadpole-pinched scheme). Lower: The chosen renormalization scheme on the left (right) for $\delta v_S$ is the $\mathrm{ZEM}_{h_2}$ scheme and for $\delta \alpha$ the OS tadpole-pinched scheme ($p^*$ tadpole-pinched scheme).}
\label{fig:ewplusqcdcorrections}
\end{figure}

We now turn to the discussion of the electroweak and QCD corrections and their impact on the relic density. In Fig.~\ref{fig:ewplusqcdcorrections}, we show the relative QCD plus electroweak corrections $\kappa_\Omega$ to the relic density as a function of the DM mass $m_A$ for all parameter points that pass our constraints. The color code indicates the dominant annihilation channel, given by $b\bar{b}$ (blue), $W^+W^-$ (purple), $h_2h_2$ (red), $h_1h_1$ (orange), and $h_1h_2$ (green), depending on the parameter point. The upper (lower) plots show the results when we use the ZEM$_{h_1}$ (ZEM$_{h_2}$) renormalization scheme for the singlet VEV $v_S$, where the $h_1$ ($h_2$) decay into $AA$ is used for the renormalization. The usage of the ZEM scheme allows us to cover the whole parameter region without kinematic restrictions on the decay used for renormalization. For $\alpha$ we apply the OS tadpole-pinched ($p^\star$ tadpole-pinched) scheme in the left (right) plots. \s

In the region, where the $b\bar{b}$ decays dominate we have corrections between 50\% and 125\%. The pure electroweak corrections (not shown here) to the relic density range between -2\% and 22\%. In terms of relative corrections to the TAC, we have for the QCD plus electroweak corrections $-33\% \ge \kappa_\sigma \ge 56\%$ and for the electroweak corrections alone $-2\% \ge \kappa_\sigma \ge -18\%$. Note that the density of points is significantly reduced compared to Fig.~\ref{fig:relicqcd}. This is due to the fact, that for some parameter points, the sum of the relative QCD and electroweak corrections to the TAC was found to be negative and exceeding 100\%, so that the NLO-corrected TAC becomes negative. We discarded these points as unphysical. \s

Once the $W$ boson threshold is reached, the QCD corrections become subdominant, and the values plotted in Fig.~\ref{fig:ewplusqcdcorrections} basically reflect the electroweak corrections. 
The $\kappa_\Omega$ values for the parameter points where the $WW$ channel dominates range between $-29$\% and +85\% for most of the parameter points. In terms of the TAC these values correspond to maximal corrections of
$41\% \ge \kappa_\sigma \ge -46\%$.
There are a few outliers, however, in the low $m_A$ range around $60...120$~GeV where we can reach up to 125\% correction, corresponding to $\kappa_\sigma = -56\%$.  \s

The electroweak corrections to the Higgs pair final states can be even larger. For the points, where these channels dominate the relic density, we have corrections between -75\% and 166\%, corresponding to $300\% \ge \kappa_\sigma \ge -62\%$. \s 

Overall, we see significantly larger corrections for the ZEM$_{h_1}$ than for the ZEM$_{h_2}$ scheme. Our investigation showed that the large corrections are due to the counterterm for $\delta v_S$ and occur for large mass differences between the decaying particle ($h_{1,2}$) and the final state particle $A$ in the decay used for the renormalization of $v_S$. Since for our parameter points the $h_1$ mass is 125~GeV and $m_{h_2} > 125$~GeV, for the points above the $W$ boson threshold larger mass differences occur between $h_1$ and $A$ than between $h_2$ and $A$, explaining the observed behaviour. \s

\begin{figure}[t]\centering\includegraphics[width=0.49\textwidth]{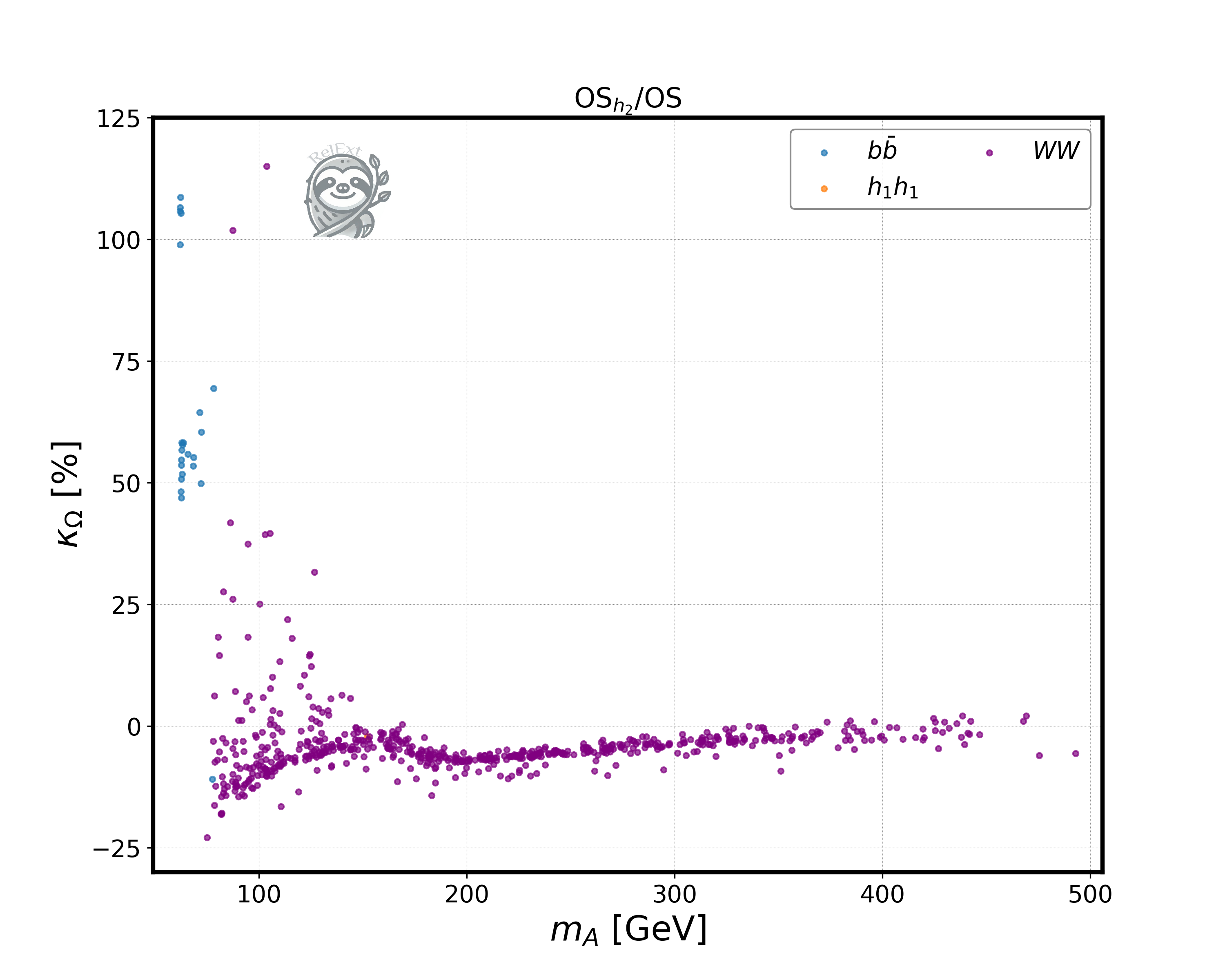}\includegraphics[width=0.49\textwidth]{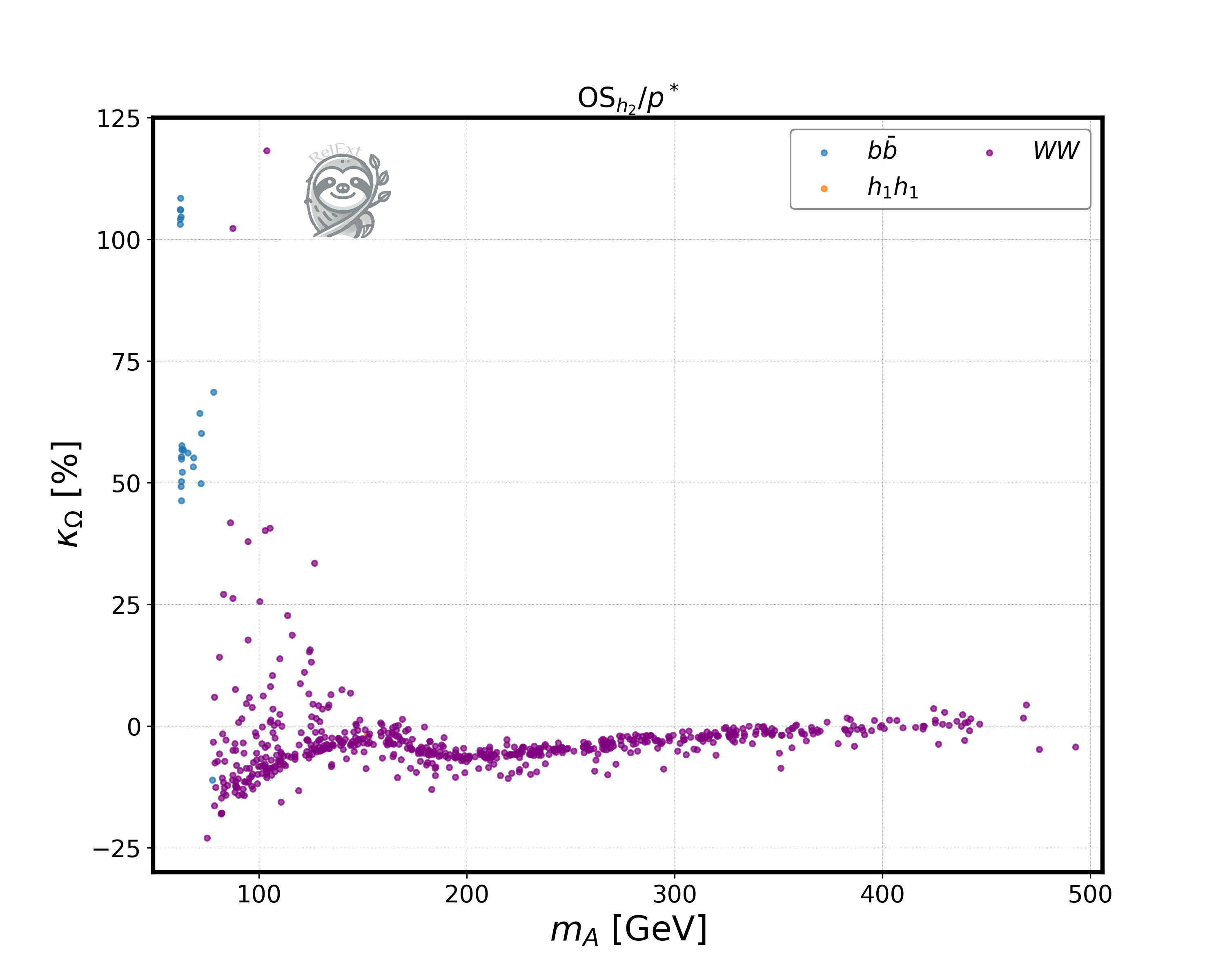}
    \caption{Relative QCD plus EW corrections $\kappa_\Omega$ to the relic density as a function of the DM mass $m_A$.  The input parameter $v_S$ is renormalized in the $\mathrm{OS}_{h_2}$ scheme and $\alpha$ is renormalized in the OS tadpole-pinched scheme (left) and the $p_*$ tadpole-pinched scheme (right). The color code indicates the dominant annihilation channel, given by  $b\bar{b}$ (blue), $W^+W^-$ (purple), and $h_1h_1$ (orange).}
\label{fig:OSSchemescxsmzemh2}
\end{figure}

We also computed the corrections by applying the process-dependent scheme for $v_S$ including the external mass momenta in the decay. This means that we cannot provide corrections for all parameter points, as for some points the kinematics will not allow for $h_{1,2}$ decays into $AA$. We discard these points. Furthermore, since we only consider the parameter space where $m_A \geq m_{h_1}/2$, we only  compute the relative corrections in the $\text{OS}_{h_2}$ scheme. The results for these points are shown in Fig.~\ref{fig:OSSchemescxsmzemh2}, where we display the relative correction $\kappa_\Omega$ on the relic density as a function of the DM mass $m_A$ with $\delta v_S$ renormalized in the  $\mathrm{OS}_{h_2}$ scheme and the mixing angle $\alpha$ renormalized in the OS tadpole-pinched (left) and the $p_\star$ tadpole-pinched scheme (right). 
The color code indicates the dominant annihilation channel, where due to our kinematic restriction of $m_{h_2} \ge 2 m_A$, the $h_2 h_2$ channel does not appear any more.
We find that the EW corrections are considerably smaller than in the ZEM scheme with $\kappa_\Omega$ values between $-23$\% and $+118$\% at most, corresponding to $30\% \ge \kappa_\sigma \ge -54\%$. The largest corrections in the low $m_A$ mass region are due to the QCD corrections. For $m_A \le 100$~GeV, the pure EW corrections are found to be at most between about $-25$\% and $100$\%. 
In the higher $m_A$ mass region where QCD corrections do not play a role any more the EW corrections are of moderate size.
While in this renormalization scheme we have a better convergence of the NLO corrections, we pay the price of not being able to apply the scheme for all parameter points. These findings are in accordance with previous findings, cf.~Ref.~\cite{Egle:2023pbm}. \s

Other reasons for large relative corrections are large quartic couplings between two DM and two Higgs particles, and hence only affect the $AA\to h_i h_j$ ($i,j=1,2$) channels. Last but not least large relative corrections can be simply due to small tree-level TACs. 
We finally note that in the mass region above the $W$ boson threshold we discarded points that have negative TACs after the inclusion of the QCD and EW corrections. Such  negative TACs were only found for the EW but not for the QCD corrections. These points require the inclusion of higher-order corrections beyond NLO, which is out of the scope of the current paper. \s

We have seen that the results for the corrections show a strong dependence on the renormalization scheme for $v_S$. We now discuss the results for different renormalization schemes of the mixing angle $\alpha$, by comparing the two left with the two right plots in Fig.~\ref{fig:ewplusqcdcorrections} and the left with the right plot in Fig.~\ref{fig:OSSchemescxsmzemh2}. By eye, there is barely any difference visible. We can use the variation in the renormalization scheme to derive the theoretical uncertainties on the relic density calculation due to missing higher-order corrections. For this, however, we consistently have to change the input parameters to the new scheme as well when changing the renormalization scheme. We have not done this in the production of the plots in Fig.~\ref{fig:ewplusqcdcorrections} and Fig.~\ref{fig:OSSchemescxsmzemh2}, so that they should only be taken as a rough hint on the size of the theoretical uncertainties. We will discuss the uncertainties in the next section in more detail. \s

\begin{figure}[h]\centering\includegraphics[width=0.49\textwidth]{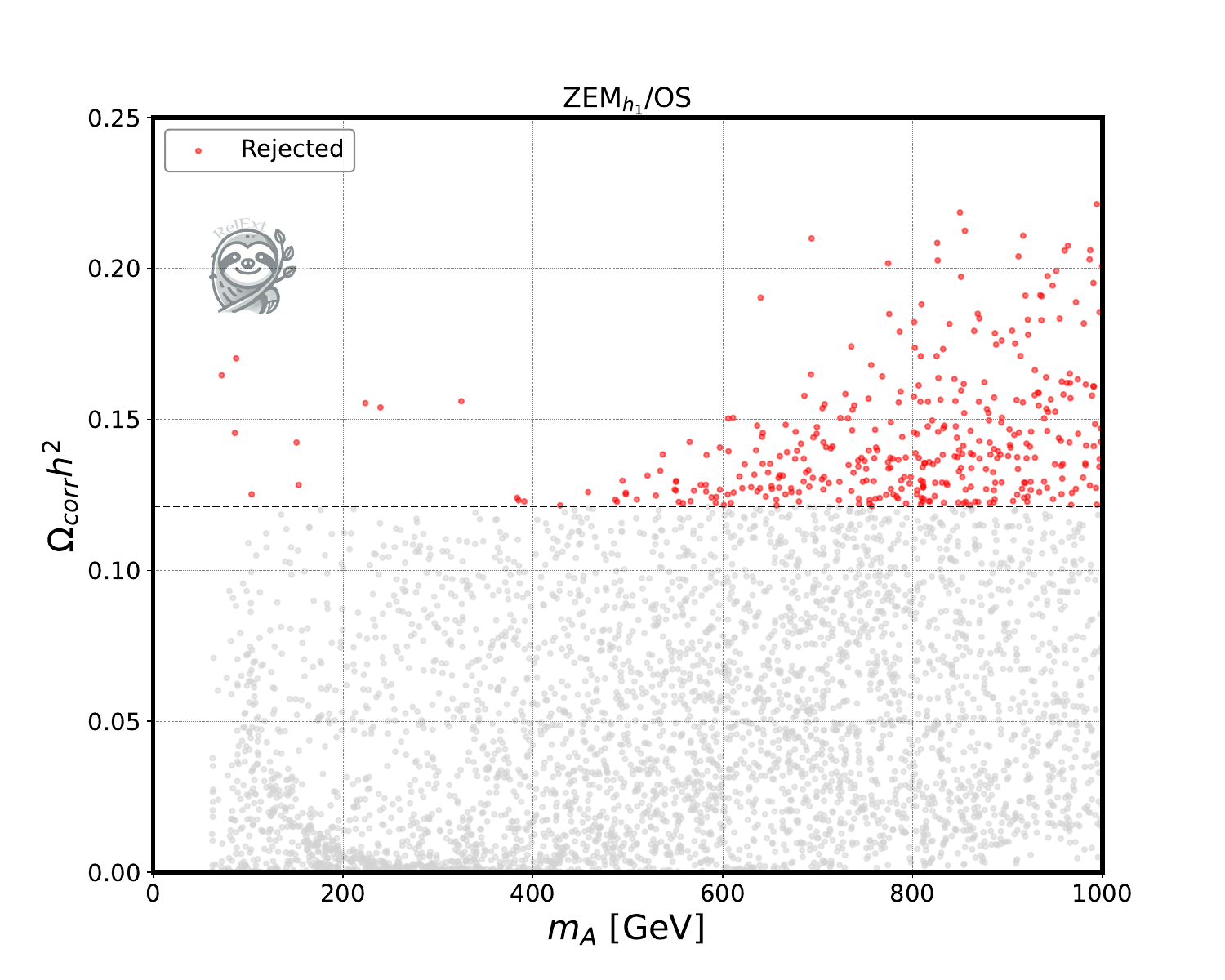}\hfill\includegraphics[width=0.49\textwidth]{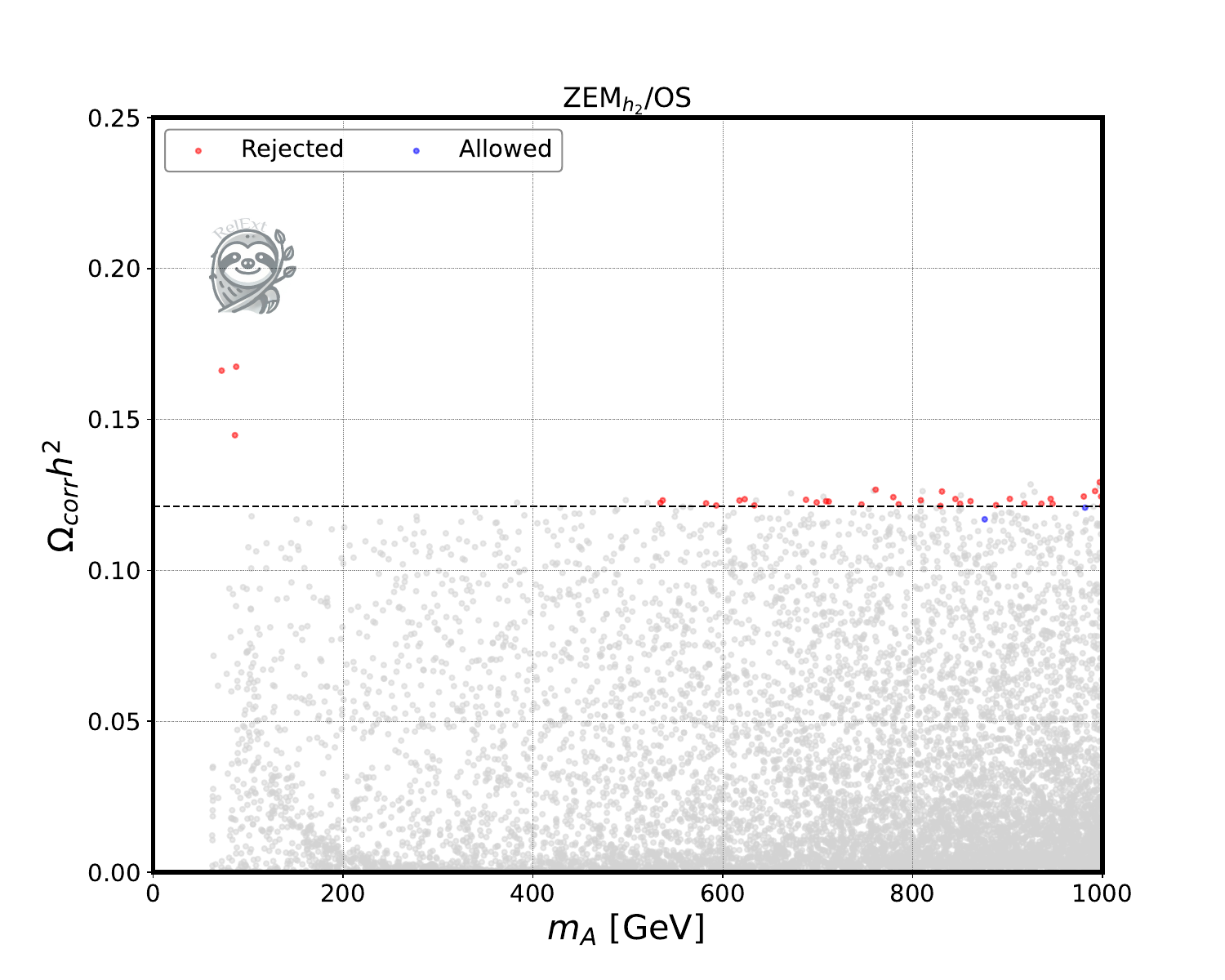}
    \caption{The QCD plus electroweak corrected relic density $\Omega_{\text{corr}} h^2$ in the ZEM$_{h_1}$ (left) and ZEM$_{h_2}$ (right) renormalization scheme for $v_S$ with $\alpha$ renormalized in the OS tadpole-pinched scheme. The dashed line indicates the measured value $\Omega h^2=0.12$. All grey points below this line are compatible with the relic density. Red (blue) points are excluded at NLO (LO) but allowed LO (NLO).}
\label{fig:reliccxsm}
\end{figure}

We finish this section by investigating the impact of the NLO corrections to the relic density on the parameter space of the model. In Fig.~\ref{fig:reliccxsm} we display the QCD plus electroweak corrected relic density $\Omega_{\text{corr}} h^2$ as a function of the DM mass $m_A$ with $v_S$ renormalized in the ZEM$_{h_1}$ (ZEM$_{h_2}$) scheme in the left (right) plot and $\alpha$ renormalized in the OS tadpole-pinched scheme. The dashed line indicates the measured value $\Omega h^2=0.12$. All grey points below this line are compatible with the relic density measurement. The red points are allowed if the LO value of the relic density is compared with the measured value, but they are excluded after the inclusion of the NLO corrections to the relic density. The blue points on the other hand are only allowed after the NLO corrections to the relic density are taken into account. At LO, they would be excluded.
This demonstrates the importance of the NLO corrections in constraining the parameter space. When relating this to the input parameters, however, we cannot identify a specific feature that would be related to this behaviour and could be targetedly tested at the LHC. This claim may change in the future, when the parameter space will get more and more constrained also from other measurements.



\subsection{Uncertainty Estimate}
The uncertainty in the computation of the relic density due to missing higher-order corrections can be estimated through a change of the renormalization scheme. We have implemented different renormalization schemes for the mixing angle and the singlet VEV. For the proper estimate of the uncertainty through the variation of the renormalization scheme, the input parameters also have to be converted to the new scheme.
We denote the renormalized parameter and its renormalization constant in the $i$th renormalization scheme by  $p_i$ and $\delta p_i$, respectively. Since the bare parameter $p_0$ does not depend on the chosen renormalization scheme, we can relate the renormalized parameters in the different schemes as
\begin{equation}
p_0= p_{1}+\ct{p}_{1}(p_{1})\overset{!}{=}p_{2}+\ct{p}_{2}(p_{2})\,.
\label{ParamConversion}
\end{equation}
If we are now interested in the renormalized parameter in scheme 2, Eq.~(\ref{ParamConversion}) is solved for $p_2$ and expanded up to one-loop order, which is the loop order we are considering here, leading to 
\begin{equation}\label{eq:paramconv}
p_{2}=p_{1}+\ct{p}_{1}(p_{1})-\ct{p}_{2}(p_{2})\approx p_{1}+\ct{p}_{1}(p_{1})-\ct{p}_{2}(p_{1})-\underbrace{\ct{p}_{2}'(p_{1})
\bigl(\ct{p}_{1}(p_{1})-\ct{p}_{2}(p_{2})\bigr)}_{\mathcal{O}(\delta p^2)}\,.
\end{equation}
In the second step we approximated $\delta p_2 (p_2)$, which would have to be solved iteratively, by $\delta p_2 (p_1)$, and consistently discarded all terms that would formally be of two-loop order. Note that the counterterm in Eq.~(\ref{eq:paramconv}) is to be understood as the one-loop counterterm. 
In order to quantize the theoretical uncertainty estimated through the change between two renormalization schemes, namely our default renormalization scheme 'default' and the chosen new renormalization scheme '$\lambda$', we introduce the relative renormalization scheme dependence $\delta_\text{ren}^\lambda$ as
\begin{eqnarray}\label{eq:uncertparam}
    \delta_\text{ren}^\lambda \equiv  \frac{\Omega^\lambda_\text{corr}h^2 - \Omega^{\text{default}}_\text{corr}h^2}{\Omega^{\text{default}}_\text{corr}h^2}
    \;,
\end{eqnarray}
where $\Omega^\lambda_\text{corr}h^2$ is the relic density calculated at NLO in the renormalization scheme $\lambda$ and $\Omega^\text{default}_\text{corr}h^2$ is the relic density at NLO in our default renormalization scheme with $v_s$ renormalized in the $\mathrm{ZEM}_{h_2}$ scheme and $\alpha$ in the OS tadpole-pinched scheme. For the consistent derivation of the uncertainty, we furthermore make sure to convert our input parameters from the initial renormalization scheme to the new scheme by applying  Eq.~\eqref{eq:paramconv}. \s

For the discussion of the uncertainty, we show results for a benchmark point, \texttt{BP1}, which is defined in Tabs.~\ref{tab:bp1} and \ref{tab:bp1_pstar}, 
respectively. We first investigate the effect of the change in the renormalization scheme of $v_S$ from the default ZEM$_{h_2}$ scheme to the new scheme $\lambda \equiv$ ZEM$_{h_1}$, cf.~Tab.~\ref{tab:bp1}. 
The required converted new input parameter $v_S$ is also given in the table. The last three columns display the LO and the QCD+EW corrected relic densities as well as the relative correction $\kappa_\Omega$ of the relic density in the ZEM$_{h_2}$ and the ZEM$_{h_1}$ scheme, 
respectively. In the latter scheme, the relic densiy is consistently obtained by using $v_S$ after conversion into the ${\text{ZEM}_{h_1}}$ scheme. We first note, that the change in the renormalization scheme leads to a visible change in the singlet VEV $v_S$ by more than 4~GeV. The change in $\kappa_\Omega$ is accordingly significant, moving from roughly 6 to 15\%. The derived theoretical uncertainty, however, amounts to only 1.17\%, with $\delta_\text{ren}^{\text{ZEM}_{h_1}}=0.0117$. This is understood by looking at the LO values for the relic densities. They change considerably due to the consistent change of the input parameter, compensating the large difference in $\kappa_\Omega$. Comparing, respectively, the two LO relic density values and the two corrected relic density values with each other, we see a considerable reduction of the uncertainty in the relic density due to the inclusion of the NLO corrections. \s

\begin{table}[h!]
\centering
\footnotesize

\begin{tabular}{c|ccccc|ccc}
    \hline\hline
Scheme & $m_{h_1}$ & $m_{h_2}$ & $m_{A}$ & $\alpha$ & $v_s$
    & $\Omega_\text{LO} h^2$ & $\Omega_\text{corr} h^2$ & $\kappa_\Omega\;[\%]$ \\
    \hline\hline
    Default ZEM$_{h_2}$ & \multirow{2}{*}{125.09} & \multirow{2}{*}{180.63} & \multirow{2}{*}{997.01}
    & \multirow{2}{*}{$-0.1293$}
    & 231.23
    & 0.1200 & 0.1268 & 5.6 \\
    New ZEM$_{h_1}$ & & & & &  
    226.99
    & 0.1119 & 0.1283 & 14.7\\
    \hline\hline
\end{tabular}
\caption{Input parameters of \texttt{BP1} in the default renormalization scheme with $v_S$ renormalized in the ZEM$_{h_2}$ and $\alpha$ renormalized in the OS tadpole-pinched scheme. The value of $v_S$ after consistent conversion into the 
$\text{ZEM}_{h_1}$ scheme via Eq.~\eqref{eq:paramconv} is given as well. The last three columns show the LO and the QCD+EW corrected relic densities as well as the relative correction $\kappa_\Omega$, for the ZEM$_{h_2}$ and the ZEM$_{h_1}$ scheme, respectively. Masses and $v_S$ are given in GeV. }
\label{tab:bp1}
\end{table}

We next investigate the effect when changing the renormalization of the mixing angle $\alpha$ from the default OS tadpole-pinched to the $p_\star$ tadpole-pinched scheme. Since the counterterm $\ct{v_S}$ depends on the mixing angle, $v_S$ must be adjusted accordingly upon a change in the renormalization of $\alpha$. Looking at Tab.~\ref{tab:bp1_pstar}, we first note that the input parameter $\alpha$, consistently obtained from Eq.~\eqref{eq:paramconv}, barely changes when we move from the OS to the $p_\star$ tadpole-pinched scheme. Also the change in $v_S$ is minimal. Accordingly, the LO relic density shows very little renormalization scheme dependence, which is further reduced after inclusion of the QCD+EW corrections. We find a theoretical uncertainty due to missing higher-order corrections of $\delta^{p_\star}_{\text{ren}} = 1.6 \times 10^{-4}$, i.e.~about 0.02\%. 

\begin{table}[h!]
    \centering
    \footnotesize
%
%
    \begin{tabular}{c|ccccc|ccc}
        \hline\hline
Scheme & $m_{h_1}$ & $m_{h_2}$ & $m_{A}$ & $\alpha$ & $v_s$
        & $\Omega_\text{LO} h^2$ & $\Omega_\text{corr} h^2$ & $\kappa_\Omega\;[\%]$ \\
        \hline\hline
Default OS &        
        \multirow{2}{*}{125.09} & \multirow{2}{*}{180.63} & \multirow{2}{*}{997.01}
        & -0.1293 & 231.23
        & 0.12001 & 0.12680  & 5.6310   \\
New $p_\star$ & & & & -0.1290 & 231.24
        & 0.12003 & 0.12678 & 5.6307 \\
        \hline\hline
    \end{tabular}    
\caption{Input parameters of \texttt{BP1} in the default renormalization scheme with $v_S$ renormalized in the ZEM$_{h_2}$ and $\alpha$ renormalized in the OS tadpole-pinched scheme. The value of $\alpha$ after consistent conversion into the 
$p_\star$ tadpole-pinched scheme via Eq.~\eqref{eq:paramconv} and of the consequently also changed $v_S$ value are given as well. The last three columns show the LO and the QCD+EW corrected relic densities as well as the relative correction $\kappa_\Omega$, for the OS tadpole-pinched and the $p_\star$ tadpole-pinched scheme, respectively. Masses and $v_S$ are given in GeV.}
    \label{tab:bp1_pstar}
\end{table}

\begin{figure}[t!]\centering\includegraphics[width=0.45\textwidth]{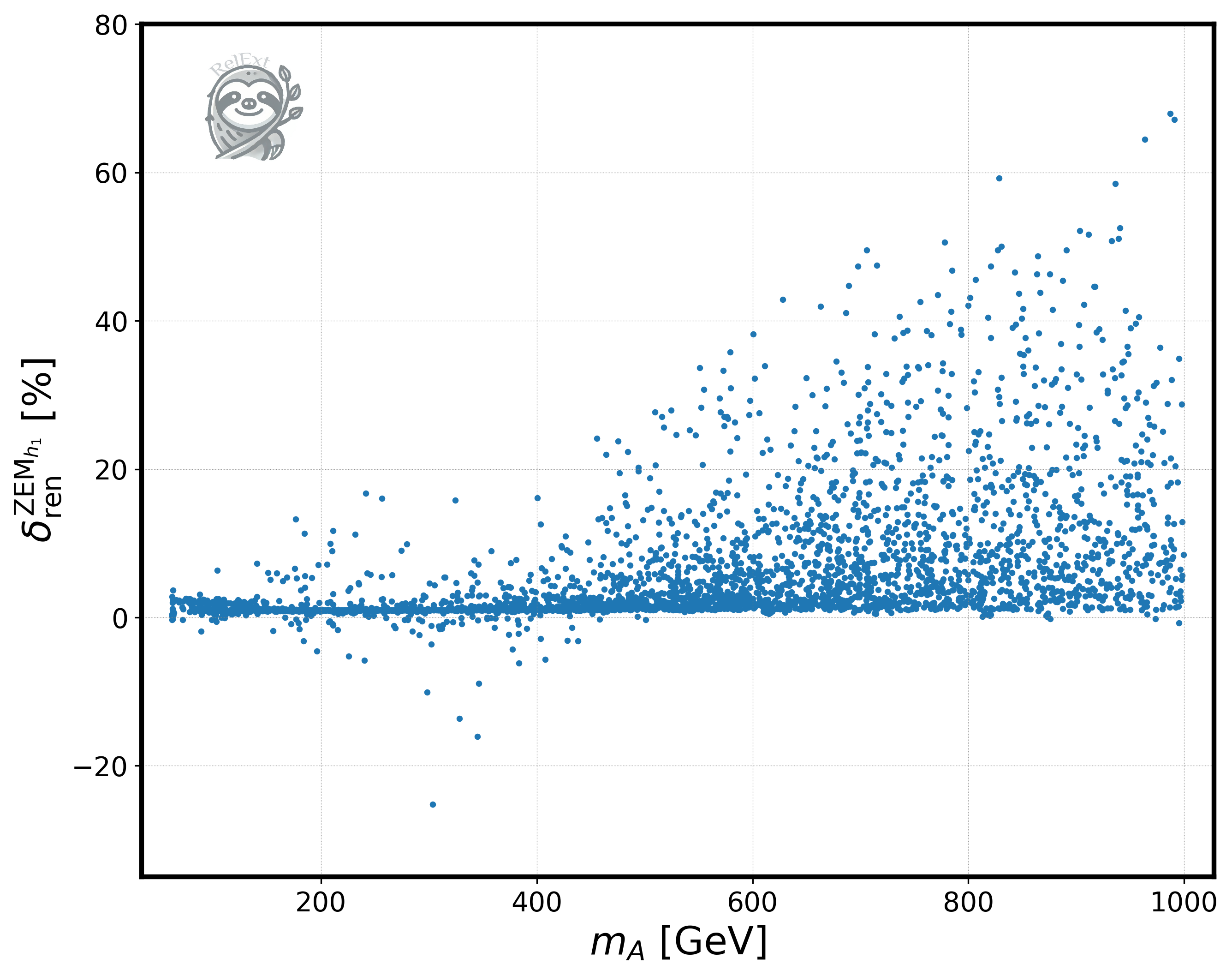}
\includegraphics[width=0.45\textwidth]{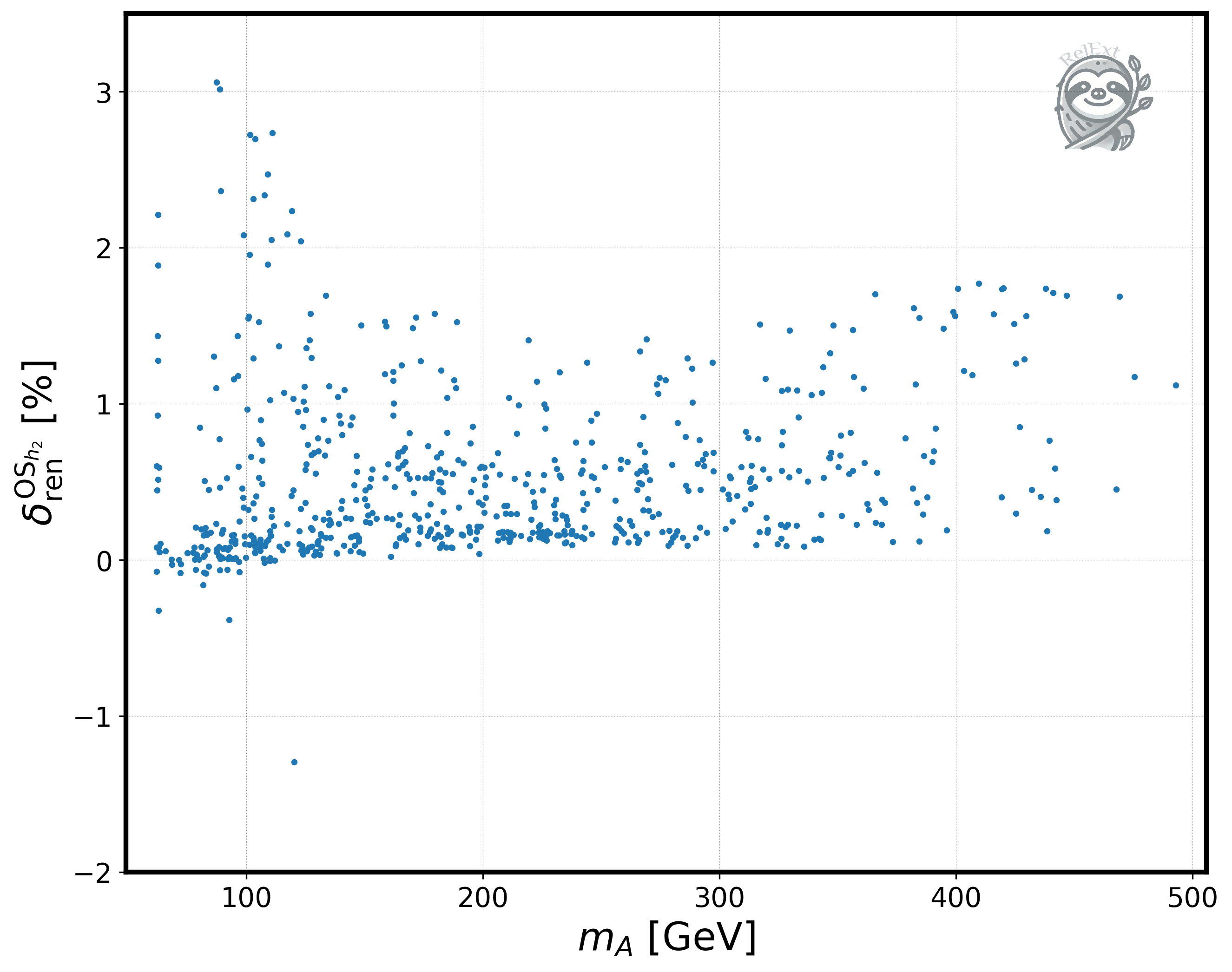}
    \caption{Estimated uncertainty $\delta_{\text{ren}}^{\mathrm{ZEM}_{h_1}}$ ($\delta_{\text{ren}}^{\mathrm{OS}_{h_2}}$) of the higher-order corrections to the relic density through the renormalization scheme change of $v_S$ from the default scheme $\text{ZEM}_{h_2}$ to $\text{ZEM}_{h_1}$ (left) and $\text{OS}_{h_2}$ (right) as a function of the DM mass.
     Note, that we used in the right plot only parameter points, which fulfill the condition $m_{h_2} > 2m_A$.}
\label{fig:deltacxsmschemes}
\end{figure}

To estimate the uncertainty on the relic density due to missing higher-order corrections through the renormalization scheme dependence across the parameter space, we calculated for the generated sample of valid parameter points the QCD+EW corrections in the default renormalization scheme with $v_S$ renormalized in the $\mathrm{ZEM}_{h_2}$ scheme and $\alpha$ in the OS tadpole-pinched scheme and in the renormalization scheme where we changed the $v_S$ renormalization scheme to the $\mathrm{ZEM}_{h_1}$ and the OS$_{h_2}$ scheme, respectively, keeping the renormalization of the mixing angle unchanged. We made sure to consistently change the input parameters according to Eq.~(\ref{eq:paramconv}) when moving from one $v_S$ renormalization scheme to the other. The corresponding  estimated uncertainties are shown in Fig.~\ref{fig:deltacxsmschemes},   $\delta^{\mathrm{ZEM}_{h_1}}_{\text{ren}}$ in the left plot and $\delta^{\mathrm{OS}_{h_2}}_{\text{ren}}$ in the right plot, as function of the DM mass $m_A$. The point density in the right plot is much smaller, as we can only consider parameter points, where $m_{h_2} > 2 m_A$. As can be inferred from the left plot the uncertainty in the lower mass range, $m_A \lsim 450$~GeV ranges between about -25 and +17\%. For larger DM masses, however, the found maximum quickly increases to about 68\%. This is in accordance with our findings on the relative corrections $\kappa_\Omega$ to the relic density in the 
$\mathrm{ZEM}_{h_1}$
renormalization scheme, displayed in Fig.~\ref{fig:ewplusqcdcorrections} (upper), where we found large corrections in particular in the large $m_A$ regime, where the annihilation into Higgs pair final states dominates the TAC. With increasing $m_A$ the difference between the decaying particle $h_1$, with $m_{h_1} < 125$~GeV in our scenarios, and $m_A$ increases, which leads to an increasing counterterm $\delta v_S$ with worse perturbative convergence and hence larger theoretical uncertainty due to missing higher-order corrections. A better convergence is achieved in the ZEM$_{h_2}$ scheme, cf.~Figs.~\ref{fig:ewplusqcdcorrections} (lower), and in the OS scheme, which uses OS decays for the renormalization. Here, no large mass gaps are possible within our scan ranges. Therefore, the estimated uncertainty based on the change to the OS$_{h_2}$ scheme, $\delta^{\mathrm{OS}_{h_2}}_{\text{ren}}$, shown in the right plot, is much smaller with values between at most -1.3 and 3.1\%. We did not vary the renormalization scheme of the mixing angle, as here the change in the results and hence the related uncertainties, are negligible as the comparison of the left and the right plots in Fig.~\ref{fig:ewplusqcdcorrections} shows. The discussion in this section leads to the conclusion that for a better perturbative convergence and hence smaller uncertainties due to missing higher-order corrections, for our parameter sample with $m_{h_1} = 125$~GeV, the $v_S$ renormalization schemes based on the $h_2$ decay into the DM pair are the preferred choice. 


\section{Conclusions}
\label{sec:conclusions}
In this paper we have provided the NLO QCD plus EW corrections to the relic density in the complex singlet extended SM, the CxSM. For this, we calculated the NLO QCD and EW corrections to the thermally averaged cross section for DM annihilation into SM particle pairs and BSM Higgs pairs. For the CxSM-specific parameters, the singlet VEV $v_S$ and the mixing angle $\alpha$, we provided the possibility to choose between different renormalization schemes. The singlet VEV is renormalized through the decays of Higgs bosons into DM pairs, where the initial Higgs states and the treatment of the external particle momenta determine the specific renormalization schemes. The mixing angle $\alpha$ is renormalized in the OS tadpole-pinched or the $p^\star$ tadpole-pinched scheme. To treat the infrared divergences, that occur in the EW corrections from loop diagrams with massless particles in the loop, we apply the Catani-Seymour dipole subtraction formalism. Our corrections are implemented in the existing tool \texttt{RelExt}. The new code \texttt{RelExt}@\texttt{NLO} can be downloaded from the url:

\centerline{\url{https://github.com/jplotnikov99/RelExt/tree/NLO}} 
\medskip
We presented the impact of our corrections for a set of parameter points obtained from a scan accross the parameter space of the CxSM that have been selected to fulfill all relevant theoretical and experimental constraints. The QCD corrections are dominating for DM masses below the $WW$ boson threshold, where the annihilation into a bottom quark pair is dominating. They are of typical size, increasing the relic density by up about 60\% depending on the parameter point. The EW corrections take over for DM masses above the $WW$ boson threshold. For scenarios where the annihiliaton into Higgs pairs is the dominating channel in the TAC, the corrections can become quite large, in particular in the high DM mass regime and for $v_S$ renormalizaton schemes based on a Higgs decay into DM pairs, where there is a large mass gap between the mass of the decaying Higgs boson and the DM pair. For our scenarios, this means a bad perturbative convergence in the ZEM$_{h_1}$ scheme, where the $h_1 \to AA$ decay is used and the external momenta are set to zero. The ZEM$_{h_2}$ and in general the OS schemes, not setting the external momenta to zero, are better. The reason is that the kinematic restrictions in combination with the chosen scan ranges prevent scenarios with large mass gaps between decaying and final state particles. The remaining theoretical uncertainties due to missing higher-order corrections estimated from the renormalization scheme change then turn out to be rather small, after the proper conversion of the input parameters. This is in line with the fact that the EW corrections are then of moderate size, ranging between -25 and +25\% for the bulk of the parameter points. We have also some points with up to 50\% for the relative correction to the relic density. Note, however, that this is induced by corrections on the TAC of -33\%, which is not unusually large for EW corrections. The few points with even larger corrections are due to large mass gaps between the decaying $h_2$ and the DM mass. As for the renormalization of the mixing angle $\alpha$, here the dependence on the renormalization scheme is negligible. \s

The here presented corrections have phenomenological impact. With the NLO corrections being positive or negative depending on the parameter values, points that are allowed by the LO relic density are excluded when the NLO corrections to the TAC and hence the relic density are included. Vice versa, points that are allowed by the NLO corrected relic density, may be excluded at LO, so that points might erroneously be discarded, if the check for compatibility is done only with the LO relic density. \s

With the here presented NLO calculation to the DM relic density in the CxSM we have reduced the theoretical uncertainty on the prediction of the relic density. The implementation of various renormalization schemes not only allows us to derive the remaining theoretical uncertainty due to missing higher-order corrections, their analysis also gives us guidelines for the development of renormalization schemes with good perturbative convergence. The next steps that we will take is the automatization of the NLO calculation to the largest possible extent, such that \texttt{RelExt}@\texttt{NLO} will be applicable for any model with a thermal DM candidate in a theory stabilized by a discrete $\mathbb{Z}_2$ symmetry. Combined with our scanning tool \texttt{ScannerS}, this will allow for the exploration of the parameter space of Higgs portal models at high precision also in the DM observables. 

\section*{Acknowledgements}
We thank Lisa Biermann for reading the draft. We thank Lisa Biermann and Duarte Azevedo for their contributions during the early stages of this work. We thank Christoph Borschensky for providing us with his interface between \texttt{LoopTools} and \texttt{Collier}, which is based on the original implementation by Julien Baglio and Matthias Kesenheimer in the \texttt{POWHEG-BOX-V2} process \texttt{weakinos-jet}. PB is grateful to the Studienstiftung des Deutschen Volkes for financial support. KE acknowledges financial support from the Avicenna-Studienwerk.
MM thanks the Deutsche Forschungsgemeinschaft (DFG, German Research Foundation) under grant 396021762 - TRR 257 for financial support. 
PG and RS acknowledge financial support from the Portuguese Foundation for Science and Technology (FCT)  under contracts: UID/PRR2/00618/2025 (https://doi.org/10.54499/UID
/PRR2/00618/2025), UID/PRR/00618/2025 (https://doi.org/
10.54499/UID/PRR/00618/2025), and UID/00618/2025 (\url{https://doi.org/10.54499/UID/00618/2025}). PG also acknowledges financial support from FCT under the PhD grant 2022.11377.BD (\url{https://doi.org/10.54499/2022.11377.BD}).

\begin{appendix}
\section*{Appendix}
\section{How to use \texttt{Relext@NLO}}\label{app:bp}
The code \texttt{RelExt@NLO} is based on the code \texttt{RelExt} \cite{Capucha:2025iml}. In the following, we describe the new options in \texttt{RelExt@NLO} and give a sample input and output file to illustrate the usage of the code.

\subsection{New Settings}
For a more detailed description of each individual settings, we refer to Ref.~\cite{Capucha:2025iml}. In this section, we will focus on the new setting options introduced in this work. A summary of these settings is provided in Tab.~\ref{tab:NLOsettings}.
\begin{table}[htbp]
\centering
\begin{tabular}{p{0.2\textwidth} p{0.6\textwidth}}
\toprule
\ttt{Setting} & Description \\
\midrule
\ttt{NLO} & If set to \texttt{true}, NLO corrections, including both EW and QCD effects, are taken into account. In this configuration, only the final states $ff \in\{t\bar{t}, b\bar{b},\tau\bar{\tau}, ZZ, W^+W^-, h_1h_1, h_1h_2, h_2h_2\}$ are taken into account. If set to 0 or \texttt{false}, the calculation is performed at LO.\\
\ttt{QCD} & If set to \texttt{true} in addition to enabling NLO, only QCD corrections are included. In this case, annihilation processes into third-generation quarks are evaluated at NLO QCD, while all remaining processes are computed at LO.\\
\ttt{renormvs} & Specifies the renormalization scheme for the counterterm $\delta v_S$.
A value of 1 corresponds to the definition given in Eq.~\eqref{vs_os1}, while 2 corresponds to Eq.~\eqref{vs_os2}.
Setting the parameter to 3 selects the ZEM scheme with the $h_1 \rightarrow AA$ condition, given in Eq.~(\ref{vs_zem1}), while 4 corresponds to the ZEM scheme with $h_2 \rightarrow AA$, defined in Eq.~(\ref{vs_zem2}).\\
\ttt{renormalpha} & Specifies the renormalization scheme used for the counterterm $\delta \alpha$. If set to 1, the OS tadpole-pinched scheme is used, which is defined in Eq.~\eqref{alpha_os}. If set to 2, the $p^*$ tadpole-pinched scheme according to Eq. ~\eqref{alpha_pstar} is used.\\
\bottomrule
\end{tabular}
\caption{Description of the new options available in the settings block of the \ttt{main.cpp} file.}
\label{tab:NLOsettings}
\end{table}

\subsection{Example: DM Relic Density at NLO QCD}
In this section, we illustrate the usage of the code by computing the NLO QCD corrections for the benchmark point given in Tab.~\ref{tab:params}. This parameter point is provided in the example input file \texttt{cxsm\_NLO\_example.tsv} (line 81), which is included with the installation.
\begin{table}[htbp]
\centering
\small
\begin{tabular}{ccccc}
\toprule
$m_{h_1}$ & $m_{h_2}$ & $m_{h_2}$ & $v_s$ & $\alpha$ \\
\midrule
 125.09 &135.504  & 68.0931 & 954.837 & -0.14767 \\
\bottomrule
\end{tabular}
\caption{Benchmark point. 
This point is included in the file \texttt{cxsm\_NLO\_example.tsv} in line 81 . }
\label{tab:params}
\end{table}

\noindent The relevant \texttt{C++} settings for the example point are shown in Listing 1.
\begin{lstlisting}[style=cppstyle, caption={C++ settings in \texttt{main.cpp} for NLO calculations.}, label=lst:nlo_settings]
/* Change to desired settings starting from here
***********************************************
*/
static constexpr int MODE = 3;
static const VecString SAVEPARS = {"MA1", "MS1", "alpha", "svev"};
static const VecString CONSIDERCHANNELS = {};
VecString NEGLECTCHANNELS = {};
static const VecString NEGLECTPARTICLES = {"u", "d", "c", "s", "e", "mu"};
static constexpr double BEPS = 1e-6;
static constexpr double XTODAY = 1e6;
static constexpr bool FAST = true;
static constexpr bool CALCWIDTHS = false;
static constexpr bool SAVECONTRIBS = true;
static constexpr bool NLO = true;
static constexpr bool QCD = true;
static constexpr int renormvs = 4;
static constexpr int renormalpha = 1;
/*
***********************************************
Until here */
}
\end{lstlisting}
In our example, the settings $\texttt{NLO}$ and $\texttt{QCD}$ are set to $\ttt{true}$, which means that only the QCD corrections are calculated. In this case, the choice of $\ttt{renormvs}$ and $\ttt{renormalpha}$ is irrelevant, as these counterterms $\delta v_S$ and $\delta\alpha$ do not contribute at QCD NLO. The $\ttt{main}$ function is given by:
\begin{lstlisting}[style=cppstyle, caption={Example \texttt{main()}-function for the NLO relic density computation.}, label=lst:main_function]
int main(int argc, char **argv) {
    
    clock_t begin_time = clock();
    std::cout << std::setprecision(16);
printRelExtInfo(NLO);
    OptionParser parser;
    
    parser.parse(argc, argv);
    int parampoint = std::stoi(parser.get("line"));
    
    //LO 
    Main M_lo(argv, MODE, BEPS, XTODAY, FAST, CALCWIDTHS, SAVECONTRIBS, 0);
    M_lo.LoadParameters(parampoint);
    M_lo.set_channels(CONSIDERCHANNELS, NEGLECTCHANNELS, NEGLECTPARTICLES);
    
    //NLO 
    Main M_nlo(argv, MODE, BEPS, XTODAY, FAST, CALCWIDTHS, SAVECONTRIBS, NLO,QCD,renormvs,renormalpha);
    M_nlo.set_channels(CONSIDERCHANNELS, NEGLECTCHANNELS1, NEGLECTPARTICLESNLO);
    M_nlo.LoadParameters(parampoint);

    // Model Information   
    std::cout << "Model used: Complex Singlet Extension of the SM (CxSM)" << std::endl;
    std::cout << "-----------------------------------------------------" << std::endl;
    std::cout << "Initial variables" << std::endl;
    std::cout << "-----------------------------------------------------" << std::endl;
    std::cout << "MH = \t" << M_nlo.GetParameter("MH")<< std::endl;
    std::cout << "MS1 = \t" << M_nlo.GetParameter("MS1") << std::endl;
    std::cout << "MA1 = \t" << M_nlo.GetParameter("MA1") << std::endl;
    std::cout << "svev = \t" << M_nlo.GetParameter("svev") << std::endl;
    std::cout << "alpha = \t" << M_nlo.GetParameter("alpha") << std::endl;
    
    std::cout << "-----------------------------------------------------" << std::endl;
    std::cout << "Relic Density Results Leading Order" << std::endl;
    std::cout << "-----------------------------------------------------" << std::endl;
    double lo_val = M_lo.CalcRelic();
    std::cout << "Omega h^2 (LO)  = " << lo_val  << std::endl;
    double nlo_val = M_nlo.CalcRelic();
    std::cout << "-----------------------------------------------------" << std::endl;
    std::cout << "Relic Density Results Next-To-Leading Order" << std::endl;
    std::cout << " " << std::endl;
    std::cout << "Omega h^2 (NLO) = " << nlo_val << std::endl;
    double rel_corr = (nlo_val / lo_val - 1.0) * 100.0;
    std::cout <<"kappa[%]  = \t" << rel_corr << std::endl;
    std::cout << "-----------------------------------------------------" << std::endl;
      /* ---------- CSV-Output ---------- */
   
    std::string csvname = "../dataOutput/" + std::string(argv[2]);
    M_nlo.SaveToCSV(csvname, M_nlo, lo_val, nlo_val, rel_corr);

    float seconds = float(clock() - begin_time) / CLOCKS_PER_SEC;
    int minutes = int(seconds) / 60;
    float remaining_seconds = seconds - minutes * 60;

    std::cout << "Time: " << minutes << " min "
          << remaining_seconds << " s\n";

    return 0;
}

\end{lstlisting}
\noindent To execute the program, use the following command:
\begin{lstlisting}
./cxsm ../dataInput/NameOfInputFile NameOfOutputFile line=LineNumberOfParameterPoint
\end{lstlisting}
where
\begin{itemize}
    \item \texttt{NameOfInputFile} is the path to your input file (e.g., \texttt{cxsm\_NLO\_example.tsv}),
    \item \texttt{NameOfOutputFile} is the desired name of the output file,
    \item \texttt{LineNumberOfParameterPoint} is the line number in the input file containing the parameter point of interest (e.g., \texttt{81}).
\end{itemize}

\noindent The program generates an output file in CSV format containing the following columns:
\begin{table}[htbp]
\centering
\small
\begin{tabular}{ll}
\toprule
Column & Description \\
\midrule
\texttt{MH} & Mass of the Higgs boson $h_1$ in GeV \\
\texttt{MS1} & Mass of the Higgs boson $h_2$ in GeV \\
\texttt{MA1} & Mass of the DM candidate $A$ in GeV \\
\texttt{svev} & VEV of the singlet field $v_S$ in GeV \\
\texttt{alpha} & Mixing angle $\alpha$ \\
\texttt{Omega\_LO} & Relic density $\Omega h^2$ at LO \\
\texttt{Omega\_NLO} & Relic density $\Omega h^2$ at NLO \\
\texttt{kappa[\%]} & Relative correction $\kappa_{\Omega}$ in \% \\
\bottomrule
\end{tabular}
\caption{Structure of the output file generated by the program.}
\label{tab:output_structure}
\end{table}

The example output file is shown below:
\begin{lstlisting}[style=csvstyle, label=lst:output_example]
MH,MS1,MA1,svev,alpha,Omega_LO,Omega_NLO,kappa[%]
125.09,135.503635414378,68.093091075082,954.836591497416,-0.147674337790616,0.06560469706851459,0.1039006181510157,58.37374882244579
\end{lstlisting}


\end{appendix}

\vspace{1ex}

\appendix

\newpage
\bibliographystyle{h-physrev.bst}
\bibliography{ref.bib}

\end{document}